\documentclass[twocolumn]{aastex62}

\usepackage{array}
\usepackage{amsmath}
\usepackage[counterclockwise, figuresleft]{rotating}
\usepackage{txfonts}
\usepackage{booktabs}
\usepackage{multirow}
\usepackage[caption=false]{subfig}
\usepackage{longtable}
\usepackage{lineno}

\newcommand{\msun}{M_\odot}
\newcommand{\rsun}{R_{\rm \odot}}
\newcommand{\lsun}{L_\odot}

\newcommand{\logg}{\log g}

\def\m2s2{\hbox{\,m$^{2}$\,s$^{-2}$}} 
\def\kms{\hbox{\,km\,s$^{-1}$}}       

\begin{document}


\title{Independent validation of the temperate Super-Earth HD79211 b using HARPS-N}


\author[0000-0003-0741-7661]{Victoria DiTomasso}\thanks{National Science Foundation Graduate Research Fellow\\Corresponding author: Victoria DiTomasso\\victoria.ditomasso@cfa.harvard.edu}

\author[0000-0001-8838-3883]{Chantanelle Nava}

\author[0000-0003-3204-8183]{Mercedes López-Morales}

\author[0000-0001-6637-5401]{Allyson Bieryla}

\author[0000-0001-5383-9393]{Ryan Cloutier}

\affiliation{Center for Astrophysics $|$ Harvard ${\rm \&}$ Smithsonian, 60 Garden Street, Cambridge, MA 02138, USA}

\author[0000-0002-6492-2085]{Luca Malavolta}

\affiliation{Dipartimento di Fisica e Astronomia “Galileo Galilei”, Universitá di Padova, Vicolo del l’Osservatorio 3, I-35122 Padova, Italy}

\author[0000-0001-7254-4363]{Annelies Mortier}

\affiliation{KICC \& Astrophysics Group, Cavendish Laboratory, University of Cambridge, J.J. Thomson Avenue, Cambridge CB3 0HE, UK}
\affiliation{School of Physics \& Astronomy, University of Birmingham, Edgbaston, Birmingham, B15 2TT, UK}

\author[0000-0003-1605-5666]{Lars A. Buchhave}
\affiliation{DTU Space, National Space Institute, Technical University of Denmark, Elektrovej 328, DK-2800 Kgs. Lyngby, Denmark}

\author[0000-0002-3481-9052]{Keivan G. Stassun}
\affiliation{Department of Physics and Astronomy, Vanderbilt University, Nashville, TN 37235, USA}

\author[0000-0002-7504-365X]{Alessandro	Sozzetti}
\affiliation{INAF - Osservatorio Astrofisico di Torino, Via Osservaorio 20, I10025 Pino Torinese, Italy}

\author[0000-0002-6177-198X]{Aldo Stefano Bonomo}
\affiliation{INAF - Osservatorio Astrofisico di Torino, Via Osservaorio 20, I10025 Pino Torinese, Italy}

\author[0000-0002-9003-484X]{David Charbonneau}
\affiliation{Center for Astrophysics | Harvard and Smithsonian, 60 Garden Street, Cambridge, MA 02138, USA}

\author{Andrew	Collier Cameron}
\affiliation{Centre for Exoplanet Science, SUPA School of Physics and Astronomy, University of St Andrews, North Haugh, St Andrews KY16 9SS, UK}

\author[0000-0003-1784-1431]{Rosario Cosentino}
\affiliation{Fundación Galileo Galilei - INAF, Rambla J. A. F. Perez, 7, E-38712 S.C. Tenerife, Spain}

\author[0000-0001-9984-4278]{Mario Damasso}
\affiliation{INAF - Osservatorio Astrofisico di Torino, Via Osservaorio 20, I10025 Pino Torinese, Italy}

\author[0000-0002-9332-2011]{Xavier Dumusque}
\affiliation{Department of Astronomy, University of Geneva, Chemin Pegasi 51, Versoix, Switzerland}

\author[0000-0002-4272-4272]{A. F. Martínez Fiorenzano}
\affiliation{Fundación Galileo Galilei - INAF, Rambla J. A. F. Perez, 7, E-38712 S.C. Tenerife, Spain}

\author[0000-0003-4702-5152]{Adriano Ghedina}
\affiliation{Fundación Galileo Galilei - INAF, Rambla J. A. F. Perez, 7, E-38712 S.C. Tenerife, Spain}

\author{Avet Harutyunyan}
\affiliation{Fundación Galileo Galilei - INAF, Rambla J. A. F. Perez, 7, E-38712 S.C. Tenerife, Spain}

\author[0000-0001-9140-3574]{R. D. Haywood}
\affiliation{Astrophysics Group, University of Exeter, Exeter EX4 2QL, UK}
\affiliation{STFC Ernest Rutherford Fellow}

\author[0000-0001-9911-7388]{David Latham}
\affiliation{Center for Astrophysics | Harvard and Smithsonian, 60 Garden Street, Cambridge, MA 02138, USA}

\author[0000-0002-1742-7735]{Emilio Molinari}
\affiliation{INAF - Osservatorio Astronomico di Cagliari, Via della Scienza 5, I-09047, Selargius, Italy}

\author{Francesco A. Pepe}
\affiliation{Department of Astronomy, University of Geneva, Chemin Pegasi 51, Versoix, Switzerland}

\author[0000-0002-4445-1845]{Matteo Pinamonti}
\affiliation{INAF - Osservatorio Astrofisico di Torino, Via Osservaorio 20, I10025 Pino Torinese, Italy}

\author[0000-0003-1200-0473]{Ennio Poretti}
\affiliation{Fundación Galileo Galilei - INAF, Rambla J. A. F. Perez, 7, E-38712 S.C. Tenerife, Spain}
\affiliation{INAF – Osservatorio Astronomico di Brera, via E. Bianchi 46, I-23807 Merate (LC), Italy}

\author[0000-0002-6379-9185]{Ken Rice}
\affiliation{SUPA, Institute for Astronomy, University of Edinburgh, Blackford Hill, Edinburgh EH9 3HJ, Scotland, UK}
\affiliation{Centre for Exoplanet Science, University of Edinburgh, Edinburgh EH93FD, UK}

\author[0000-0001-7014-1771]{Dimitar Sasselov}
\affiliation{Center for Astrophysics $|$ Harvard ${\rm \&}$ Smithsonian, 60 Garden Street, Cambridge, MA 02138, USA}

\author[0000-0003-0996-6402]{Manu Stalport}
\affiliation{Observatoire de Geneva, Universite de Geneve, 51 ch. des Maillettes, CH-1290 Sauverny, Switzerland}

\author[0000-0001-7576-6236]{Stéphane Udry}
\affiliation{Department of Astronomy, University of Geneva, Chemin Pegasi 51, Versoix, Switzerland}

\author[0000-0002-9718-3266]{Christopher Watson}
\affiliation{Astrophysics Research Centre, School of Mathematics and Physics, Queen’s University Belfast, Belfast, BT7 1NN, UK}

\author[0000-0001-8749-1962]{Thomas G. Wilson}
\affiliation{Centre for Exoplanet Science, SUPA School of Physics and Astronomy, University of St Andrews, North Haugh, St Andrews KY16 9SS, UK}






\begin{abstract}

{We present high-precision radial velocities (RVs) from the HARPS-N spectrograph for HD79210 and HD79211, two M0V members of a gravitationally-bound binary system. We detect a planet candidate with a period of $24.421^{+0.016}_{-0.017}$ days around HD79211 in these HARPS-N RVs, validating the planet candidate originally identified in CARMENES RV data alone. Using HARPS-N, CARMENES and HIRES RVs spanning a total of 25 years, we further refine the planet candidate parameters to $P=24.422\pm0.014$ days, $K=3.19\pm0.27$ m/s, $M$ sin $i = 10.6 \pm 1.2 M_\oplus$, and $a = 0.142 \pm0.005$ au. We do not find any additional planet candidate signals in the data of HD79211 nor do we find any planet candidate signals in HD79210. This system adds to the number of exoplanets detected in binaries with M dwarf members, and serves as a case study for planet formation in stellar binaries.}

\smallskip
\smallskip
\smallskip
\smallskip
\smallskip

\end{abstract}

\section{Introduction}\label{sect:intro}

M dwarfs are the most common stars, constituting as many as 75\% of the stars in our galaxy \citep{henry_solar_2006}, and every M dwarf is predicted to host at least one planet \citep{dressing_occurrence_2015}. The abundance of M dwarfs and the prevalence of planets around them, combined with the relatively large signals planets impose on low-mass stars, have made M dwarfs the primary target for many exoplanet studies. Previous studies report that between 23\% and 46\% of M dwarfs are in binaries \citep{ward-duong_m-dwarfs_2015,winters_solar_2019,susemiehl_orbital_2021}, but so far only about 30 exoplanets (less than 7\% of the total found around M-dwarfs) have been found around binaries where at least one member is an M dwarf. Approximately 20 planets have been found orbiting M-dwarf components in an S-type configuration, characterized by the planet orbiting around just one star in a binary, around M-dwarf components \citep{thebault_planet_2015}, and less than 10 circumbinary planets have been found around binaries with an M dwarf member \citep[e.g.][]{doyle_kepler-16_2011, orosz_kepler-47_2012, schwamb_planet_2013, jain_indication_2017}. Hundreds of exoplanets have been, and will continue to be, discovered around M dwarfs, allowing for studies of planet formation in binaries with M dwarf members.


HD79210 and HD79211 are a pair of M0V stars that comprise a gravitationally-bound binary. These two stars are in many ways, twins. They are nearly identical in mass, radius
and effective temperature according to literature measurements (see Table \ref{tab:starpars_old}), varying by only a few percent at
most. Together they form one of only ten stellar binary exoplanet host systems with a binary mass ratio greater than 0.9. With a period of $1295 \pm 180$ years and semi-major axis of $130.9 \pm 5.1$ au \citep{gonzalez-alvarez_carmenes_2020}, these stars have a projected separation of 108.54 au and a semi-major axis of 130 au \citep{gonzalez-alvarez_carmenes_2020}. Many studies of disk evolution as well as exoplanet formation and occurrence around binary stars cite 100 au either projected separation or semi-major axis as a boundary for various behaviors. For example: binaries with separations greater than 100 au do not seem to affect planet populations, in comparison to single stars \citep{desidera_properties_2007,mugrauer_multiplicity_2009,kraus_impact_2016}; closer binaries ($<100$ au separation) are less likely to host planets \citep{roell_extrasolar_2012,bergfors_stellar_2013,kraus_impact_2016}; and planets in wide binaries (100 au $< a <$ 700 au) tend to have inclinations that are aligned with those of the stellar binary \citep{christian_possible_2022}. With only around $4\%$ of all exoplanet-hosting stars being members of binary or multiple systems \citep{marzari_planets_2019}, the HD79210/HD79211 binary system serves as a valuable edge-case data point for these binary planet population studies.

While no transiting planet has yet been detected around either of these targets, \cite{gonzalez-alvarez_carmenes_2020} published an RV detection of a planet candidate around HD79211 using high-precision radial velocities from the CARMENES spectrograph \citep{quirrenbach_carmenes_2016}. They reported a planet with an orbital period of $24.45\pm0.02$ days, a semi-major axis of $0.141\pm0.005$ au, a semi-amplitude of $3.07\pm0.37$ m/s, minimum mass of $10.27^{+1.47}_{-1.38}$ or $9.97^{+1.47}_{-1.38} M_\oplus$, depending on their adopted stellar mass. They found one significant periodic signal in the RV data of HD79210 at 16.3 days, which they attributed to the stellar rotation period.

In this paper, we search for the 24.4 day planet around HD79211 in the HARPS-N data taken as part of the HARPS-N GTO Rocky Planet Search (RPS) program, as well as HIRES data from \cite{butler_lces_2017}, in an effort to support this planet's presence and refine its parameters. We also look for additional signals in the HARPS-N, HIRES and CARMENES data of both targets. In section \ref{sect:data}, we discuss the data used throughout this paper. In section \ref{sect:starchar}, we discuss our efforts to charactize the two stars. In section \ref{sect:LC_analysis}, we share our analysis of the new photometric light curves taken with KeplerCam. In section \ref{sect:RVmethods}, we present our radial velocity (RV) analysis, and in section \ref{sect:discussion} we report our results and discussion.


\section{Data}\label{sect:data}

\subsection{HARPS-N Spectroscopy and White Light Radial Velocities}\label{sect:HN_spec}

These two targets were observed using the HARPS-N spectrograph installed on the 3.6 m Telescopio Nazionale Galileo (TNG) at the Observatorio del Roque de los Muchachos in La Palma, Spain \citep{cosentino_harps-n_2012, cosentino_harps-n_2014}. HARPS-N is an updated version of HARPS at the ESO 3.6 m telescope \citep{mayor_setting_2003}, with a proven RV stability better than 1 m/s.

\begin{figure}
\centering
\includegraphics[width=0.45\textwidth]{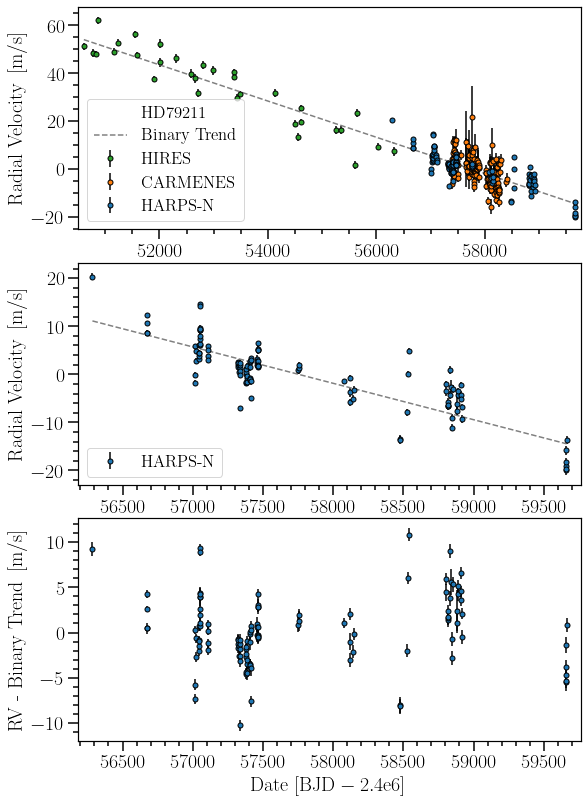}
\centering
\caption{{Radial velocities for HD79211 measured from CARMENES (orange, top panel only), HARPS-N (blue, all panels) and HIRES (green, top panel only) spectra. For most data points, the uncertainty in the RV is smaller than the size of the point. We mark the binary trend, fit as a second-order polynomial, with a dashed line in the top and middle panels. The bottom panel shows the HARPS-N radial velocities with the binary trend subtracted. We use the HARPS-N RVs with the binary trend removed when calculating the correlation between RV and activity indicators (Figure \ref{fig:act_inds_11}), and when calculating periodograms (Figures \ref{fig:pgram_RVonly_11}, \ref{fig:HN_RV_pgram_11}, \ref{fig:HN_RV_pgram_alias}, \ref{fig:stacked_pgram_HN_RV}).}}
\label{fig:HN_RVs_11}
\end{figure}

There are a total of 81 observations of HD 79210 taken between January 2014 and March 2022 (Figure \ref{fig:HN_RVs_10}). There are a total of 114 observations of HD 79211 taken between December 2012 and March 2022 (Figure \ref{fig:HN_RVs_11}). Both objects were observed as part of the HARPS-N guaranteed time observation (GTO) Rocky Planet Search program. The aim of this program is to monitor the radial velocities of nearby, bright, quiet stars in search of low-mass planets orbiting them \citep{motalebi_harps-n_2015}. For this program, targets were observed with 15-minute exposure times to minimize the effect of stellar noise with short typical time scales, i.e. averaging out p-mode oscillations \citep{chaplin_filtering_2019}. Additional measurements of the targets are spaced by two or five hours during the night to minimize the effect of granulation \citep{pepe_harps_2011,dumusque_planetary_2011,motalebi_harps-n_2015}.

These spectra were reduced with version 3.7 of the HARPS-N Data Reduction Software \citep[DRS,][]{pepe_harps_2011}. This pipeline calculates the spectral bisector inverse slope (BIS), full width at half maximum (FWHM) and contrast, as well as each observation's H$\alpha$, Na-Index, and Mount Wilson S-index $S_{MW}$. We included all of the above values in our analysis as stellar activity indicators.

We calculated radial velocities from these spectra using the HARPS-TERRA software, which calculates radial velocities using a spectral template derived from the observations themselves, and has been shown to be particularly effective when applied to M dwarfs (see \cite{anglada-escude_harps-terra_2012} for more information). The resulting RV measurements have median uncertainties of $0.65$ m/s for HD79210 and $0.62$ m/s for HD79211.

\subsection{HARPS-N Chromatic Radial Velocities}\label{sect:HN_chrom_RVs}

We also used the HARPS-N spectra to calculate chromatic radial velocities for HD79211 to test the potential effect of stellar activity. We divided the spectra into three wavelength ranges, 383.0-446.89 nm, 443.26-513.87 nm and 510.39-690 nm, and then calculated the RV of the target based on each spectral range independently. This procedure is explained in detail in \cite{mortier_k2-111_2020} and summarized here. We coadded all of the CCFs calculated using version 3.7 of the HARPS-N Data Reduction Software (DRS) from each Echelle order that belonged to the desired wavelength range, and performed a Gaussian fit on the result in order to obtain the corresponding chromatic RV. We estimated the RV error from the photon noise, the total counts of the coadded CCFs, and the read-out noise. We applied a correction factor to the noise estimate in order to closely reproduce the RV error from the full-spectrum CCF, knowing that the chromatic RVs will inevitably be noisier than the full-spectrum RVs. The average error on the full-spectrum HARPS-N DRS RVs for both targets was 0.56 m/s, and the median error on the chromatic RVs determined by each spectral range was 2.35 m/s, 1.14 m/s and 0.99 m/s, respectively. Note that these chromatic radial velocities were calculated using the CCFs from the HARPS-N DRS, not using the HARPS-TERRA software like the white light RVs used in the rest of this work.


\subsection{Absolute Astrometry}

The Gaia EDR3 archive reports two statistical indicators of possible deviations from a single-star solution that might indicate the presence of orbiting companions, \texttt{astrometric\_excess\_noise} and RUWE (renormalized unit weight error). The values of the two parameters are 0.127 mas, 1.08 and 0.138 mas, 1.11 mas, for HD 79210 and HD79211, respectively. The values of \texttt{astrometric\_excess\_noise} are in line with those obtained for very bright stars, for which the calibration of Gaia astrometry will require further improvements. The RUWE values are clearly below the empirical threshold of 1.4 above which the single-star solution is deemed unsatisfactory \citep{lindegren_gaia_2018,lindegren_gaia_2021}, and which might be an indication of the presence of an unresolved companion. A more stringent threshold at RUWE $\gtrsim 1.1$ has been proposed in recent studies \citep{belokurov_unresolved_2020,stassun_parallax_2021} for the identification of unresolved stellar systems, but this is better used for selection of samples rather than a single source-basis. The historical RV and astrometric time-series (see \citealt{gonzalez-alvarez_carmenes_2020}, and references therein) do not show obvious evidence of additional, massive unresolved companions orbiting either HD 79210 or HD 79211. We finally note that astrometric acceleration catalogues such as those from \citet{kervella_stellar_2022} and \citet{brandt_hipparcosgaia_2021} show statistically significant (signal-to-noise ratio $>$ 10) proper motion anomalies at the mean Gaia epoch for both components. This information can be used to further improve the constraints on the binary orbit, but such a study goes beyond the scope of this paper.

\subsection{Previously Published High-Precision Radial Velocities}

There are 70 high-precision RV measurements of HD79210 and 159 of HD79211 published in \cite{gonzalez-alvarez_carmenes_2020}.
These measurements were obtained with the VIS channel of the CARMENES spectrograph on the 3.5m telescope of the Calar Alto Observatory \citep{quirrenbach_carmenes_2016}. The observations were taken from March 2016 to January 2019 and January 2016 to October 2018 respectively, with median uncertainties of 2.2 and 2.0 m/s respectively. \cite{gonzalez-alvarez_carmenes_2020} also published NIR RVs, but with uncertainties averaging 9 m/s, they did not have high enough precision to search for the $\approx3$ m/s signal at 24.4 days.

We also include 32 radial velocities of HD79210 that were obtained between June 1997 and December 2013 with a median uncertainty of 1.27 m/s, and 32 radial velocities of HD79211 that were obtained between June 1997 and February 2013 with a median uncertainty of 1.53 m/s using the HIRES spectrograph \citep{vogt_hires_1994} on the Keck telescope \citep{butler_lces_2017}. With only one to a few data points per season, these RVs are not sufficient to constrain any planet candidate orbital parameters on their own, but are used in our analysis, where we include them as additional points alongside the HARPS-N and CARMENES radial velocities.

\smallskip
\smallskip
\smallskip

\subsection{Photometric Light Curves}\label{sect:LC_data}

HD79210 and HD79211 are bright M dwarfs in a binary, and are separated by 17 arcseconds on the sky. Because of this small angular separation, these two targets are blended in many photometric surveys, such as ASAS-SN \citep{shappee_man_2014,kochanek_all-sky_2017}, WASP \citep{pollacco_wasp_2006} and TESS \citep{ricker_transiting_2015}. \cite{gonzalez-alvarez_carmenes_2020} did obtain optical photometric timeseries of the two stars, but lacking adequate reference stars, calculated their differential photometry using each member of the binary as a reference for the other. Ultimately, they were not able to attribute any photometric variability to one star or the other. Because of this, we collected new light curves for these two targets. 


We obtained new photometric light curves for both HD79210 and HD79211, with the aim of determining their stellar rotation periods. We took images using the KeplerCam wide-field CCD camera on the 1.2 m telescope at the Fred Lawrence Whipple Observatory (FLWO) on Mount Hopkins, Arizona. The 4096 X 4096, 15-micron pixels format provides a 23.1 arcmin square  field of view with a resolution of 0.338 arcsec per pixel \citep{szentgyorgyi_keplercam_2005} and our images were binned by two.

From February until June 2021, we took a total of 6081 images of HD79210 and HD79211 using the B-band filter on KeplerCam. The exposure time of each image was 2 seconds. We reduced the raw images, first by stitching together the four individual images from each quadrant of the detector \citep{carter_transit_2011} and then performing bias subtraction and flat-field correction. We performed photometry using AstroImageJ (AIJ) \citep{collins_astroimagej_2017}. We used AIJ's variable radius, multi-aperature photometry function, by which AIJ determines the aperture radius from an azimuthally averaged radial profile, centered on the user-defined aperture. It then determines the net integrated counts within the object aperature, as well as the average value of the pixels in the user-defined sky annulus, not including background pixels that are more than two sigma above the average background pixel value \citep{collins_astroimagej_2017}, and returns the difference between those values as the Source-Sky counts. AIJ also calculates an error value on the Source-Sky counts, which includes contributions from read-out noise, dark current, and source and sky Poisson noise as defined by \cite{merline_realistic_1995}.

\begin{figure}
\centering
\includegraphics[width=0.5\textwidth]{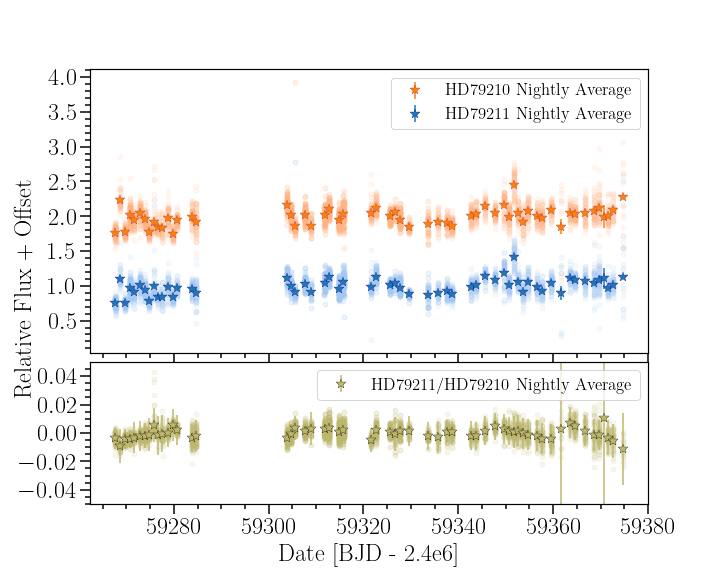}
\centering
\caption{Relative photometry for HD79210 using a comparison star (orange), relative photometry for HD79211 using a comparison star (blue), and relative photometry for HD79211 using HD79210 as a comparison star (green), plus an offset, as taken with KeplerCam. Note the different y-axis scale between the two separate light curves in the upper panel and the combined light curve in the lower panel. The faint, round points represent the relative flux measurement from each individual image. The darker, star-shaped points represent the nightly averaged relative flux. These average, nightly fluxes are what we used when calculating the periodograms (Figure \ref{fig:LC_pgram}). These two separate light curves are highly affected by systematics, which we discuss in detail in Section \ref{sect:LC_analysis}.}
\label{fig:LC}
\end{figure}

We used AIJ to calculate the Source-Sky counts and error on those measurements for our two targets, HD79210 and HD79211, as well as three comparison stars found within the field: HD233614, HD79450 and TYC 3806-1026-1. We present three sets of relative photometry: that of HD79210 calculated using TYC 3806-1026-1 as a comparison star, that of HD79211 calculated using TYC 3806-1026-1 as a comparison star, and that of HD79211 calculated using HD79210 as a comparison star (see Section \ref{sect:LC_analysis} for more information). We opted to combine the multiple exposures taken over the course of the night into nightly measurements. The resulting light curves, after sigma clipping outlier points, are shown in Figure \ref{fig:LC}.

\subsection{Literature Photometry}\label{sectsub:survey_phot}

As with the time series surveys discussed in Section \ref{sect:LC_data}, HD79210 and HD79211 are blended in many past  photometric surveys as well, due to their small separation on the sky. In some surveys, such as WISE, these two stars are blended beyond separability. We opt not to quote the magnitudes for the blended source here. In other surveys, such as GALEX, 2MASS and SDSS, the photometric values suffer from some amount of blending or diffraction spike confusion which has been corrected in post-processing, and magnitudes for these objects are reported in those survey catalogues. In even higher angular resolution surveys, namely  Hipparcos and Tycho, Gaia, and USNO, the two stars are resolved. We have listed all available photometric values in Table \ref{tab:survey_phot}, noting those that come from surveys in which the two stars were originally blended.


\section{Stellar Characterization}\label{sect:starchar}

The HD79210/HD79211 binary is located 6.332 parsecs from Earth \citep{gaia_collaboration_gaia_2020}. These two stars are very similar to one another and are in many ways twins. They are nearly identical in mass, radius and effective temperature according to literature measurements (see Table \ref{tab:starpars_old}), varying by only a few percent at most. HD79210 had been classified as a K7V star \citep{kirkpatrick_standard_1991}, but more recently has been classified as an M0V star, like HD79211 \citep{alonso-floriano_carmenes_2015}.

In Table \ref{tab:starpars_old}, we summarize the previously published parameters for these two stars. For the majority of the parameters, we cite the same sources used by \cite{gonzalez-alvarez_carmenes_2020}, with one addition. These two stars have since been included in \cite{sarmento_determination_2021}'s sample of 313 M dwarf stars, for which they derived $T_\text{eff}$ and metallicity using H-band APOGEE spectra.

For our investigation, we attempted a number of different methods to calculate the stellar parameters for the two stars in this system, namely Stellar Parameter Classification (SPC) \citep{buchhave_abundance_2012, buchhave_three_2014}, isochrone fitting \citep{morton_isochrones_2015, mortier_k2-111_2020}, empirical color-parameter relations \citep{benedict_solar_2016,boyajian_stellar_2012,mann_how_2015} and spectral energy distribution (SED) fitting \citep{stassun_eclipsing_2016}. Ultimately, we decide to use the literature values for the stellar parameters (Table \ref{tab:starpars_old}) in our analysis, for the following reasons. SPC does not produce reliable results for these targets as they are at the edge of the range of stellar effective temperatures for which SPC is calibrated. The isochrone fitting procedure takes the SPC-derived effective temperature and metallicity as inputs, as well as the flagged 2MASS photometry, making the isochrone results similarly unreliable. HD79210 and HD79211 fall at or beyond the upper range of stellar masses, depending on the true mass of the stars, for which the mass-luminosity relationship \citep{boyajian_stellar_2012} was calibrated. The color-temperature relationship \citep{mann_how_2015} also relies on the flagged 2MASS photometry.

Finally, we determined the stellar masses, radii and effective temperatures via spectral energy distribution (SED) fitting, following the procedure in \cite{stassun_eclipsing_2016}. These results were the least consistent with the other methods and the previously published values, finding that the stars are cooler, more metal-poor and have larger radii than otherwise measured. This could have an astrophysical origin: low-mass stars in much tighter, eclipsing binaries ($P<10$ days) have been shown to have larger radii and cooler temperatures than predicted by models. Current theory explains that tidal effects between the two members of the binary cause the stars to rotate quickly and be more active. Either through magnetic fields suppressing convective energy transport or high spot coverage lowering the average temperature of the stellar surface and forcing a larger radius to conserve energy, the stars end up being larger and cooler than otherwise predicted (\cite{ribas_fundamental_2008} and the references therein). These stars' wide binary separation, lack of evidence of fast rotation and other activity indicators not pointing toward them being extremely active, however, do not support this explanation of radius inflation and temperature suppression. More likely, the contamination between the stars in their photometry is causing an unreliable SED fit. For these reasons, we decide not to report the stellar parameters determined using these various methods, and opt use the literature values for the stellar parameters (Table \ref{tab:starpars_old}) in our analysis.

\begin{table*}[t]
  \caption{Stellar parameters for HD 79210 and HD 79211 found in the literature. These are the values that we ultimately use for fitting the radial velocities and deriving the parameters of the planet candidate. In the cases where multiple values are listed for a single parameter, * denotes the value adopted in this work.} \label{tab:starpars_old}
         \small
         \centering
   \begin{tabular}{lccr}
            \hline
            \noalign{\smallskip}
            Parameter  & HD 79210 & HD 79211 & Ref. \\
            \noalign{\smallskip}
            \hline
            \noalign{\smallskip}
            RA [deg] & 138.5835616152243 & 138.5913117966687 & \cite{gaia_collaboration_gaia_2016} \\
            DEC [deg] & 52.68408016691263 & 52.68342944184038 & \cite{gaia_collaboration_gaia_2016} \\
            
            Spectral type & M0V & M0V & \cite{alonso-floriano_carmenes_2015} \\
            Distance [pc] & 6.332 $\pm$ 0.002 & 6.332 $\pm$ 0.002 & \cite{gaia_collaboration_gaia_2020}  \\
            Parallax [mas] & 157.8879 $\pm$ 0.0197 & 157.8825 $\pm$ 0.0211 & \cite{gaia_collaboration_gaia_2020} \\
            $\mu_\alpha$ cos $\delta$ [mas a$^{-1}$] & -1545.787 $\pm$ 0.018 & -1573.040 $\pm$ 0.018 & \cite{gaia_collaboration_gaia_2020} \\
            $\mu_\delta$ [mas a$^{-1}$] & -569.053 $\pm$ 0.018 & -659.906 $\pm$ 0.019 & \cite{gaia_collaboration_gaia_2020} \\
            U [\kms] & -42.20 $\pm$ 0.36 & -44.01 $\pm$ 0.36 & \cite{cortes-contreras_vizier_2016} \\
            V [\kms] & -14.99 $\pm$ 0.10 & -17.44 $\pm$ 0.10 & \cite{cortes-contreras_vizier_2016}\\
            W [\kms] & -23.73 $\pm$ 0.34 & -23.10 $\pm$ 0.34 & \cite{cortes-contreras_vizier_2016} \\
            Eff. temp., T$_{\rm eff}$ [K] & 4001 $\pm$ 100* & 4014 $\pm$ 100* & \cite{sarmento_determination_2021} \\
            & 4024 $\pm$ 51 & 4005 $\pm$ 51 & \cite{schweitzer_carmenes_2019}\\
            Surface grav, $\logg$ [cgs] 
            & 4.68 $\pm$ 0.07 & 4.68 $\pm$ 0.07 & \cite{schweitzer_carmenes_2019} \\
            Metallicity, $\rm [M/H]$ [dex] & 4.71 $\pm$ 0.1 & 4.63 $\pm$ 0.1 & \cite{sarmento_determination_2021} \\
            Metallicity, $\rm [Fe/H]$ & -0.05 $\pm$ 0.16 & -0.03 $\pm$ 0.16 & \cite{schweitzer_carmenes_2019} \\
            Luminosity [$\lsun$] & 0.0789 $\pm$ 0.0038 & 0.0792 $\pm$ 0.0031 & \cite{schweitzer_carmenes_2019} \\
            Mass [$\msun$] & 0.69 $\pm$ 0.07* & 0.64 $\pm$ 0.07* & \cite{gonzalez-alvarez_carmenes_2020}  \\
            & 0.591 $\pm$ 0.047 & 0.596 $\pm$ 0.042 & \cite{schweitzer_carmenes_2019}\\
            Radius [$\rsun$] & 0.58 $\pm$ 0.02 & 0.58 $\pm$ 0.03 & \cite{schweitzer_carmenes_2019}  \\
            v sin i [km s$^{-1}$] & 2.9 $\pm$ 1.2 & 2.3 $\pm$ 1.5 & \cite{glebocki_systematic_2005, reiners_carmenes_2018} \\
            $P_{rot}$ [days] & 16.3$^{+3.5}_{-1.3}$ & 16.61 $\pm$ 0.04& \cite{gonzalez-alvarez_carmenes_2020} \\
            Age [Gyr] & 1-7 & 1-7 & \cite{gonzalez-alvarez_carmenes_2020} \\
            v$_{mic}$ [\kms] & 4.71 & 4.63 & \cite{sarmento_determination_2021}\\
            v$_{mac}$ [\kms] & -0.06 & -0.11 & \cite{sarmento_determination_2021}\\
            \noalign{\smallskip}
            \hline
     \end{tabular}  
\end{table*}


\section{Light Curve Analysis}\label{sect:LC_analysis}

We present three sets of relative photometry for these stars: that of HD79210 calculated using TYC 3801-1026-1 as a comparison star, that of HD79211 calculated using TYC 3801-1026-1 as a comparison star, and that of HD79211 calculated using HD79210 as a comparison star. There were three comparison stars in the field of our observations, all of which are at least two orders of magnitude fainter than the target stars (B magnitudes of 9.92, 9.33, 11.55 in comparison to HD79211's B magnitude of 7.966) and are different spectral types (all are F stars, as opposed to the targets, which are M0s) \citep{hog_tycho-2_2000}. We decided to use TYC 3801-1026-1 as our comparison star when calculating the separate relative photometry for HD79210 and HD79211, because it best corrected for the effect of changing airmass over the course of our observations.

There were a number of nights of observations that we opted not to use from our KeplerCam photometric measurements. The seven nights of observation between March 21, 2021 and March 29, 2021 were significantly offset in both relative and absolute flux from the rest of the nights' fluxes for both targets, likely due to moon contamination. Since we were not able to correct for this offset in the relative photometry using the available comparison stars, we decided to exclude these nights from of our analysis. We also opted not to use two other nights, April 4th and April 13th, 2021, because the seeing changed significantly over the course of the night, resulting in unreliable flux measurements. 

\begin{figure}
\centering
\includegraphics[width=0.5\textwidth]{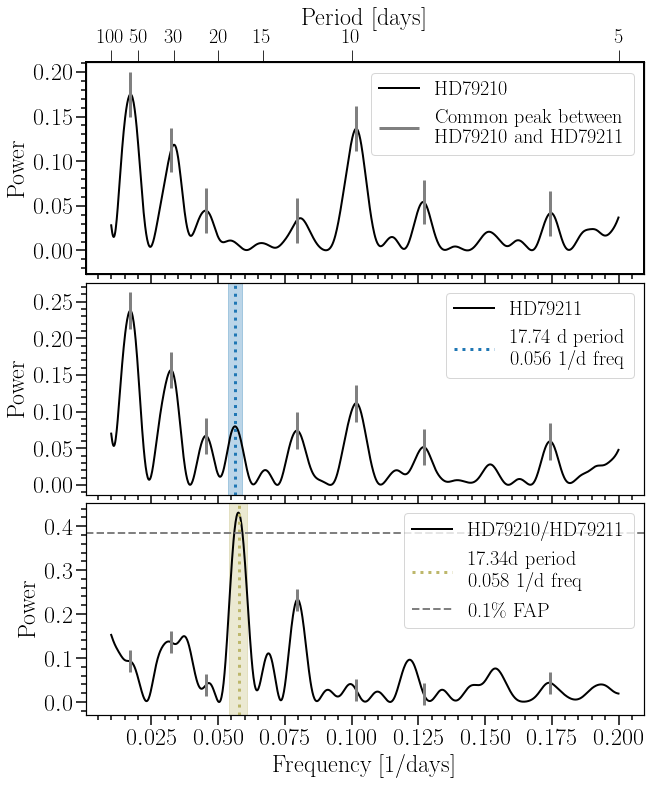}
\centering
\caption{Lomb-Scargle periodogram of the nightly-averaged relative fluxes for HD79210 using a separate comparison star (top), HD79211 using a separate comparison star (middle) and HD79211 using HD79210 as a comparison star (bottom). The vertical, grey dashes note the shared peaks between the two stars' periodograms, which are likely caused by systematics in the images. The statistically significant ($<0.1\%$ FAP) peak in the bottom periodogram is marked with the dotted green line, and the shaded region marks one standard deviation around the peak (P=17.34 days, freq $=0.057\pm0.003 \text{ days}^{-1}$). This peak corresponds to the one peak in HD79211's periodogram which is not present in that of HD79210, marked with the blue dotted line in the middle panel, at 17.74 days (freq $=0.056\pm0.003 \text{ days}^{-1}$). Although the 17.74 day peak is not statistically signficant, this suggests that this photometric variability originates in HD79211, and we attribute it to this star's rotation period.}

\smallskip
\smallskip
\smallskip
\smallskip
\smallskip
\smallskip
\smallskip
\smallskip

\label{fig:LC_pgram}
\end{figure}

The resulting light curves from the two stars (Figure \ref{fig:LC}, in orange and blue) and their periodograms (Figure \ref{fig:LC_pgram}, upper two panels) clearly show the influence of systematics. Even by eye, the time variations of these two light curves appear to be identical, indicating that they are the results of systematics affecting the images from which both stars' light curves were reduced. We rule out the far less likely scenario that the two stars were undergoing identical variation during the time of our observations. Due to the available comparison stars being bluer and fainter than our targets, these systematics persist despite using other stars in the field for calibration.

Although the other stars in the field are poorly-suited comparisons for HD79210 and HD79211, each target serves as an ideal comparison for the other, because of their similar masses, shared spectral type, and close proximity on the sky. We calculated relative photometry of HD79211 using HD79210 as a comparison (Figure \ref{fig:LC}, in green), which corrects for the systematics found in the other light curves. A periodogram of this combined light curve shows a statistically significant ($<0.1\%$ FAP) peak at a period of 17.34 days and frequency of $0.057\pm0.003 \text{ days}^{-1}$ (Figure \ref{fig:LC_pgram}, bottom panel). Alone, we cannot attribute this periodicity to one star or the other. A nearby peak, however, at 17.74 days / $0.056\pm0.003 \text{ days}^{-1}$ is present in the periodogram of HD79211's separate light curve but not in HD79210's separate light curve. Although this 17.74 day peak is not statistically significant, its presence suggests that this approximately 17.5 day periodicity originates in HD79211's photometric variability. We attribute this variability to HD79211's rotation period.

\section{Radial Velocity Analysis}\label{sect:RVmethods}

\subsection{Stellar Binary Trend Removal}\label{sect:bin_trend_removal}

The stellar binary, with its 1300 year period and nearly 1 km/s amplitude, introduces a significant RV trend to HD79210 and HD79211. {Following \citep{gonzalez-alvarez_carmenes_2020}, we tried approximating the stellar binary trend as a line, but found that a line fit did not capture the curvature introduced by extending our RV time baseline with the addition of the HIRES and HARPS-N datasets. We ultimately opted for modeling the stellar binary as a second-order polynomial, fit to the combined HARPS-N, CARMENES and HIRES data set (Figures \ref{fig:HN_RVs_11},
\ref{fig:HN_RVs_10}). In order to find correlations between the HARPS-N RVs and the various activity indicators (Section \ref{sect:RVs_and_ActInds}) and calculate periodograms of the radial velocities (Sections \ref{sect:pgrams_whitelight}, \ref{sect:pgrams_chrom}), we subtracted offsets between these three datasets and the binary polynomial trend. We used \texttt{RadVel} to fit for these offsets and binomial parameters \citep{fulton_radvel_2018}.}

{When fitting the RVs to find the planetary parameters, we chose to fit a polynomial trend (to model the binary trend), a Keplerian (to model the planet) and Gaussian Processes (GPs, to model stellar activity) simultaneously. We did this using \texttt{PyORBIT} \citep{malavolta_pyorbit_2016}, which allows the user to specify which datasets are use to fit which models -- allowing us to fit the binary trend using the combined HARPS-N, HIRES and CARMENES datasets, while varying which datasets contributed to the Keplerian and GP fits (Section \ref{sect:RVfits}).}

\subsection{HARPS-N RV Correlation with Stellar Activity Signatures}\label{sect:RVs_and_ActInds}

\begin{figure}
\centering
\includegraphics[width=0.45\textwidth]{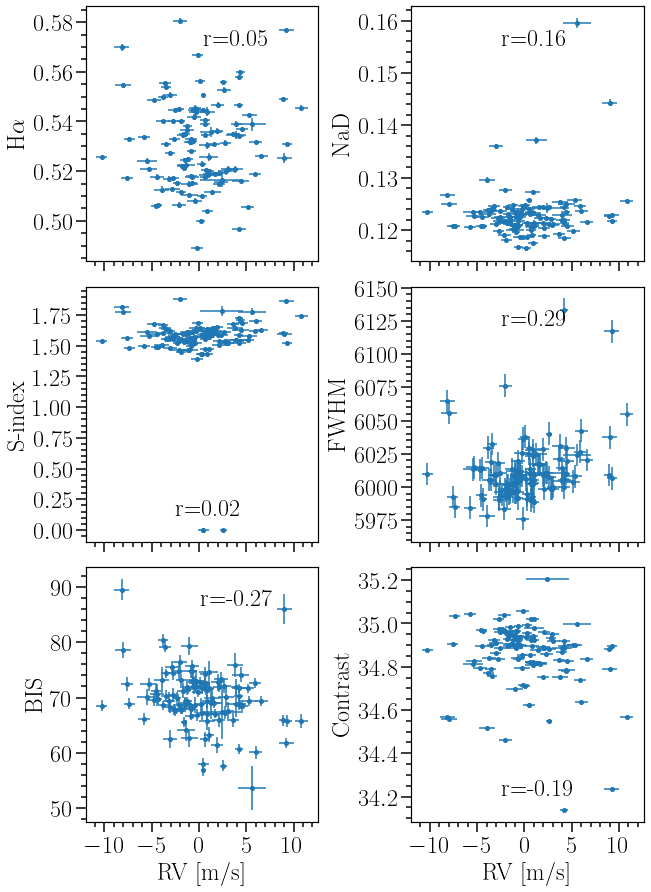}
\centering
\caption{Various HARPS-N stellar activity indicators versus HD79211's radial velocities after the stellar binary trend has been subtracted. The r-values are Pearson correlation coefficients. All of these parameters are weakly correlated, with r-values less than 0.3.}
\label{fig:act_inds_11}
\end{figure}

The HARPS-N stellar activity indicators were calculated as described in section \ref{sect:HN_spec}. We calculate the Pearson correlation coefficients ($r$) for the HARPS-N radial velocities and activity indicators using SciPy's \texttt{pearsonr} function \citep{oliphant_python_2007} for HD79210 and HD79211 (Figures \ref{fig:act_inds_10} and \ref{fig:act_inds_11}, respectively). We define a moderate correlation as $0.5>r\geq0.3$ and a weak correlation as $r<0.3$. HD79211's radial velocities are weakly correlated with all activity indicators. Radial velocities from HD79210, however, are moderately correlated with FWHM and weakly correlated with the rest.

\subsection{Periodograms of the White Light HARPS-N RVs}\label{sect:pgrams_whitelight}

\begin{figure*}
    \centering
    \includegraphics[width=0.95\textwidth]{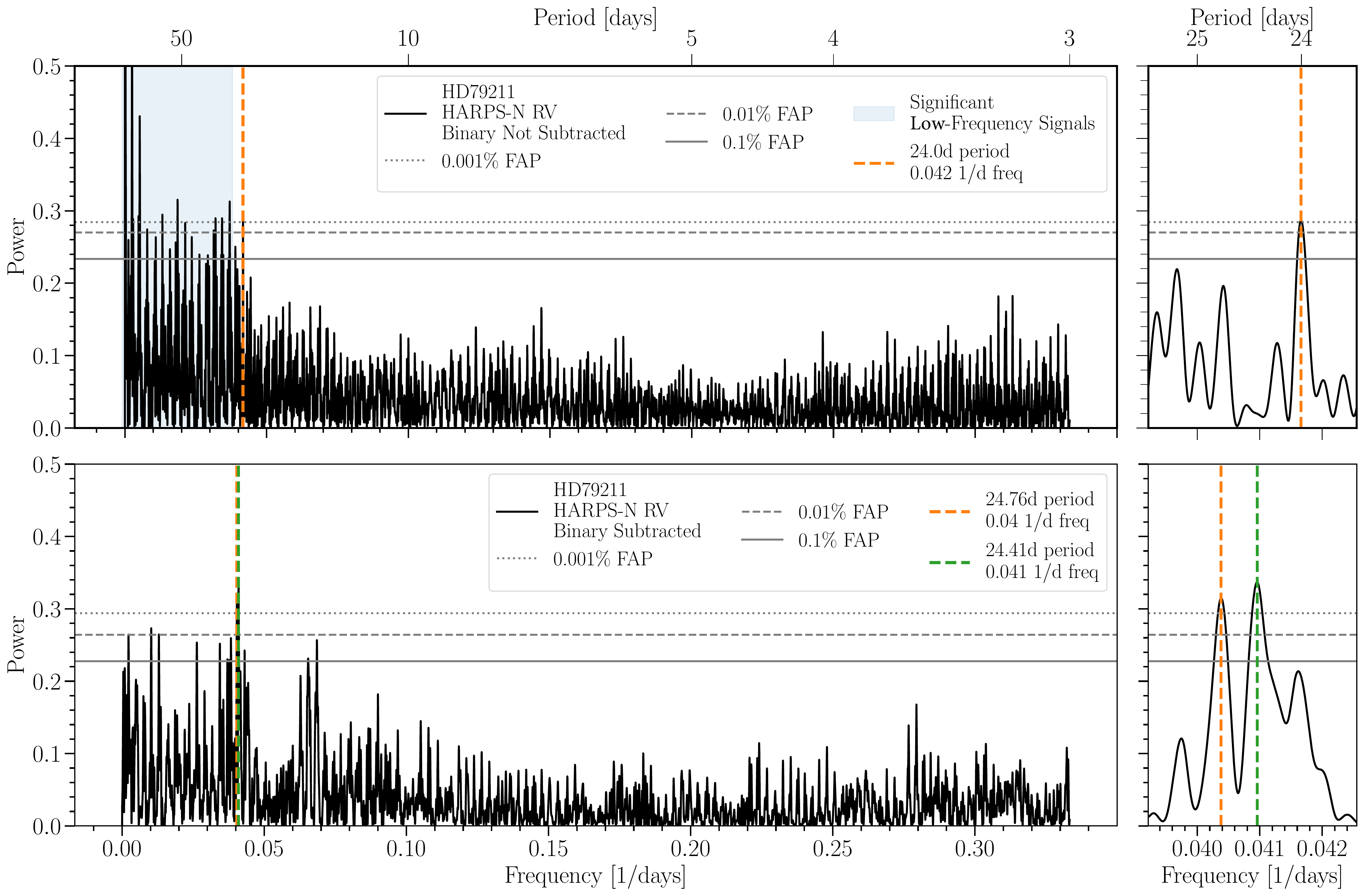}
    \caption{{Periodograms of HARPS-N radial velocities of HD79211, without the binary trend removed (top) and with the binary trend removed (bottom). Signals with frequencies from 0 to 0.33 $\text{ days}^{-1}$ are plotted. Signals with False Alarm Probabilities less than 0.1\% are marked. Before the binary trend is removed, there are many significant low frequency signals (blue shaded region), as we would expect from the long period binary. Two signals with FAPs less than 0.01\% are marked for HD79211 once the binary trend is removed, at 24.41, and 24.76 days. The signal at 24.41 days is attributed to the planet candidate, and has the highest power of the present peaks.}
    \smallskip
\smallskip
\smallskip}
    \label{fig:pgram_RVonly_11}
\end{figure*}

\begin{figure}
\centering
\includegraphics[width=0.5\textwidth]{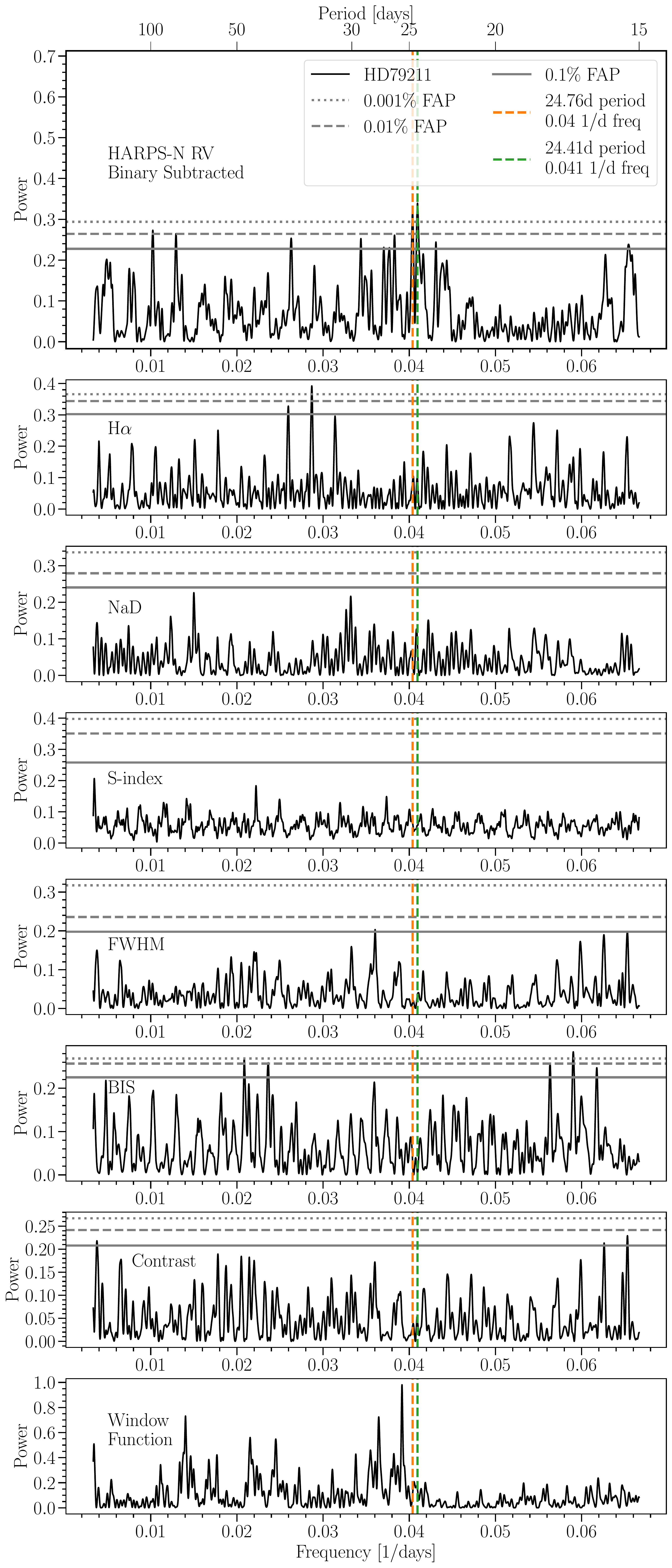}
\centering
\caption{Periodograms of HD79211's white light HARPS-N radial velocities with the binary trend removed, activity indicators, and the window function. All of significant peaks ($<0.01\%$ FAP)  are marked with dashed vertical lines. The most significant peak is at 24.41 days, which we attribute to the planet candidate. We attribute the other significant peak, at 24.76 days, to aliases of the planet candidate signal (see Figure \ref{fig:HN_RV_pgram_alias}).}
\label{fig:HN_RV_pgram_11}
\end{figure}

\begin{figure}
\centering
\includegraphics[width=0.5\textwidth]{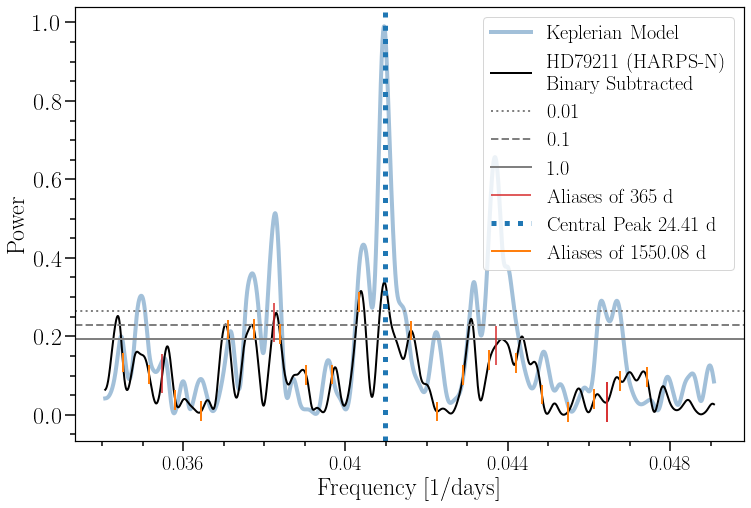}
\centering
\caption{Zoomed in periodogram of the HD79211 white light HARPS-N RVs, highlighting the strongest peak in the periodogram at 24.41 days, and the aliases that result in its interaction with the 365 day and 1550.8 day peaks in the window function. The periodogram calculated from the best fit Keplerian planet orbit, sampled with the observing cadence of the HARPS-N data, is plotted in light blue. There is variation between the periodograms of the model and the data, caused by the data's uncertainties and the influence of stellar activity in the data, but much of the aliasing structure can be seen in both periodograms. The 24.76 day peak (above the 0.01\% FAP line) can be explained as an alias.}
\label{fig:HN_RV_pgram_alias}
\end{figure}

Periodograms of the HARPS-N RVs of HD79210 and HD79211 with the binary trend removed, calculated using \texttt{Astropy}'s \texttt{LombScargle} function \citep{the_astropy_collaboration_astropy_2018}, are shown in Figures \ref{fig:pgram_RVonly_10} and \ref{fig:pgram_RVonly_11}, respectively. {The False Alarm Probabilities (FAPs) are calculated via bootstrap simulations.} The periodogram of HD79211's radial velocities reveals signals with less than 0.01\% FAP at the following periods: 24.41 and 24.76 days. The signal at 24.41 days is the most significant and we attribute it to the planet candidate. When plotted with the periodograms of the various stellar activity indicators and the window function (Figure \ref{fig:HN_RV_pgram_11}), we see that these signals in the RV periodograms do not correspond with peaks in the stellar activity indicators. We attribute the other significant peak, at 24.76 days to an alias of the 1550 day peak in the window function interacting with the 24.41 day peak (Figure \ref{fig:HN_RV_pgram_alias}). Aliases can appear in periodograms at frequencies equal to the true frequency of a signal, in this case 1/24.4 days, plus or minus the frequency in the window function caused by the data sampling, in this case 1/1550 days. We support this interpretation by plotting the periodogram of the best-fit Keplerian planetary orbit over that of the data with the binary trend removed in Figure \ref{fig:HN_RV_pgram_alias}. While the uncertainty of the data and the effects of stellar activity prevent the model and data periodograms from matching precisely, this shows that we are capable of detecting a 24.42 day signal with the observing cadence of our data, and that many of the aliases in our data correspond to those in the model. The periodogram of HD79210's RVs, after the binary trend is subtracted, does not reveal any signals with less than a 0.01\% FAP (Figure \ref{fig:pgram_RVonly_10}).

\begin{figure}
\centering
\includegraphics[width=0.5\textwidth]{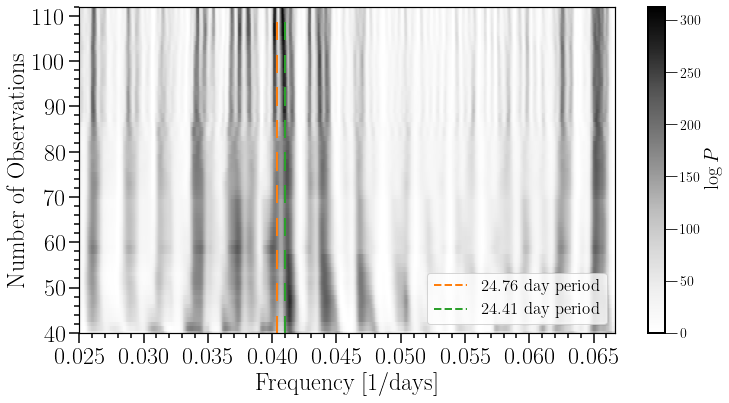}
\centering
\caption{Stacked Bayesian Generalized Lomb Scargle periodograms (SBGLS) for the HD79211 HARPS-N white light radial velocities, with the binary trend removed. The horizonal axis is the frequency of the signal, the color is the power and the vertical axis shows how the periodogram changes as observations are added to the calculation. This periodogram is oversampled by a factor of 5. The same peaks are highlighted from the 2D periodogram, Figure \ref{fig:HN_RV_pgram_11}. The strongest peak at 24.41 days gets stronger as more observations are added, as we would expect for a planet signal.}
\label{fig:stacked_pgram_HN_RV}
\end{figure}

We also present the stacked Bayesian General Lomb-Scargle periodogram (Figure \ref{fig:stacked_pgram_HN_RV}, calculated via the procedure in \cite{mortier_stacked_2017}). This stacked periodogram shows that the 24.41 day signal becomes stronger as more observations are added, as we would expect for a planet signal. We also note that the approximately 24.4 day periodic signal is present in the periodogram of HD79211's RVs regardless of how we corrected for the binary trend, whether as a line, Keplerian or as a second-order polynomial, as we are showing here. We suspect that the stellar rotation period is not present as a significant signal in these periodograms because, over the course of the observations, the magnetically active regions of the star evolved such that the stellar rotation fell out of phase.

\subsection{Periodograms of the Chromatic HARPS-N RVs}\label{sect:pgrams_chrom}

\begin{figure}
\centering
\includegraphics[width=0.5\textwidth]{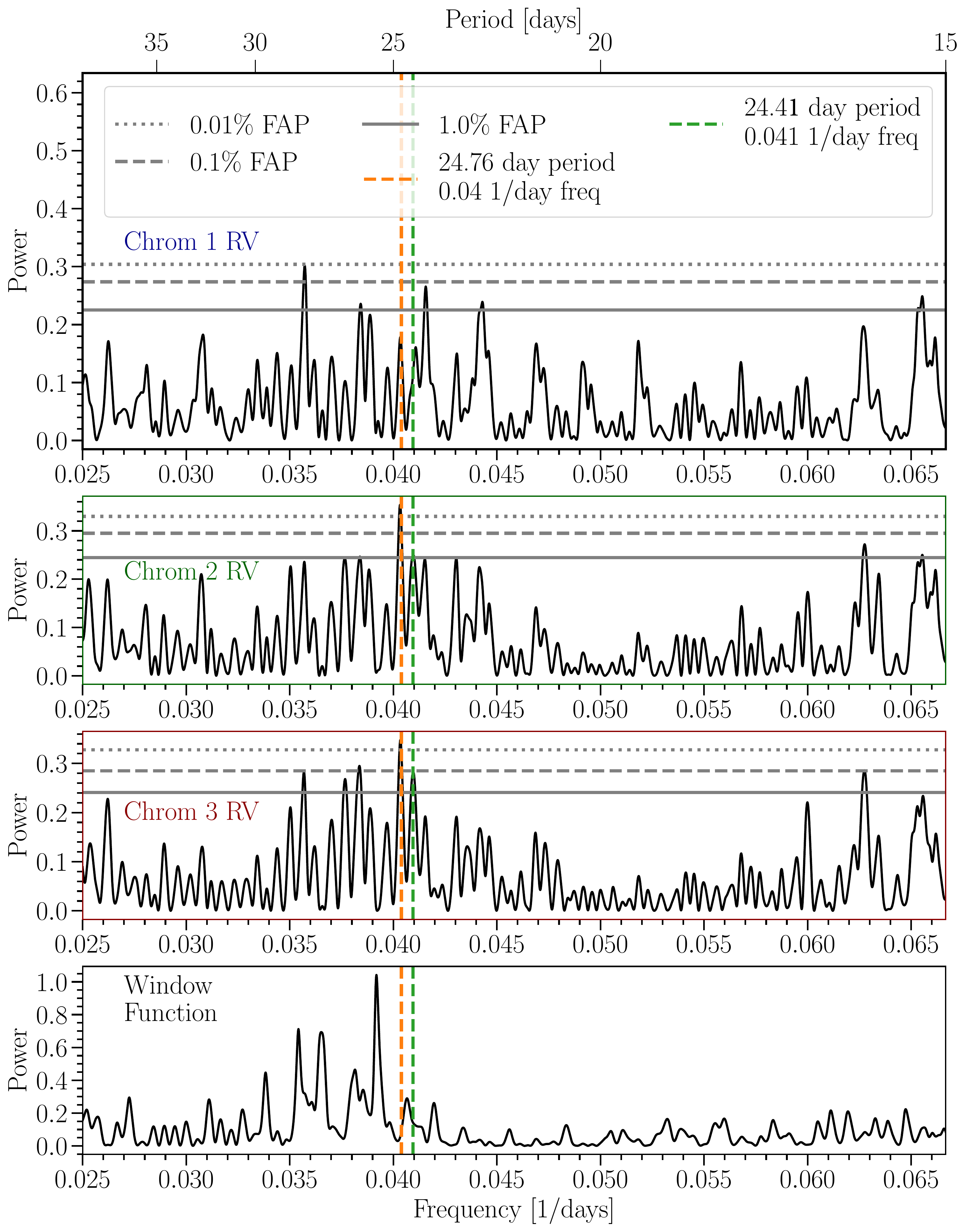}
\centering
\caption{Periodograms of the three chromatic HARPS-N RV sets and the window function. The top panel shows chrom1 RVs (calculated using the bluest portion of the spectrum), followed by chrom2, and chrom3 in the lowest RV panel (calculated using the reddest portion of the spectrum). The window function is shown in the bottom panel. The same significant peaks that are marked in the white light RV periodogram (Figure \ref{fig:HN_RV_pgram_11}) are marked here with vertical dashed lines. Both signals are present in these chromatic periodograms, with increasing significance as we look to the redder RVs. Because this signal is not larger in the bluer wavelengths, as we would expect if the signal were caused by stellar activity, we do not take this as evidence that the 24.41 day signal is caused by stellar activity.}
\label{fig:chrom_pgram}
\end{figure}

The significant peaks from the periodograms of the white light HARPS-N radial velocities are also highlighted in the periodograms of the chromatic radial velocities (Figure \ref{fig:chrom_pgram}). Both signals from the white light RVs are present in these chromatic periodograms, with increasing significance as we look to the redder RVs. Because this signal is not larger in the bluer wavelengths, as we would expect if the signal were caused by stellar activity, we do not take this as clear evidence that the 24.41 day signal is caused by stellar activity. We investigate the variation in the amplitude of the 24.41 day signal by fitting each set of chromatic radial velocities using \texttt{PyORBIT}, and discuss those results in Section \ref{sect:RVfits}.

\subsection{Radial Velocity Fits}\label{sect:RVfits}

{To determine the planetary parameters, we fit the radial velocities via an MCMC fitting routine using \texttt{PyORBIT} \citep{malavolta_pyorbit_2016}. We fit the data with a second order polynomial to model the stellar binary, a Keplerian to model the planet candidate, GPs to model stellar activity, as well as offset and jitter parameters for all of the datasets. As described in Section \ref{sect:HN_spec}, we simultaneously fit a second order polynomial (to model the binary trend) to the combined HARPS-N, CARMENES and HIRES datasets, while fitting GPs (to model the stellar activity) and a Keplerian (to model the planet) to various combinations of the three datasets. We used this procedure to fit for a planet signal in the data of HD79210 and HD79211, but were not able to identify any planet candidates in the HD79210 data. In this section, we present the resulting best-fit parameters for the planet candidate around HD79211.}

\begin{figure*}
\centering
\includegraphics[width=0.9\textwidth]{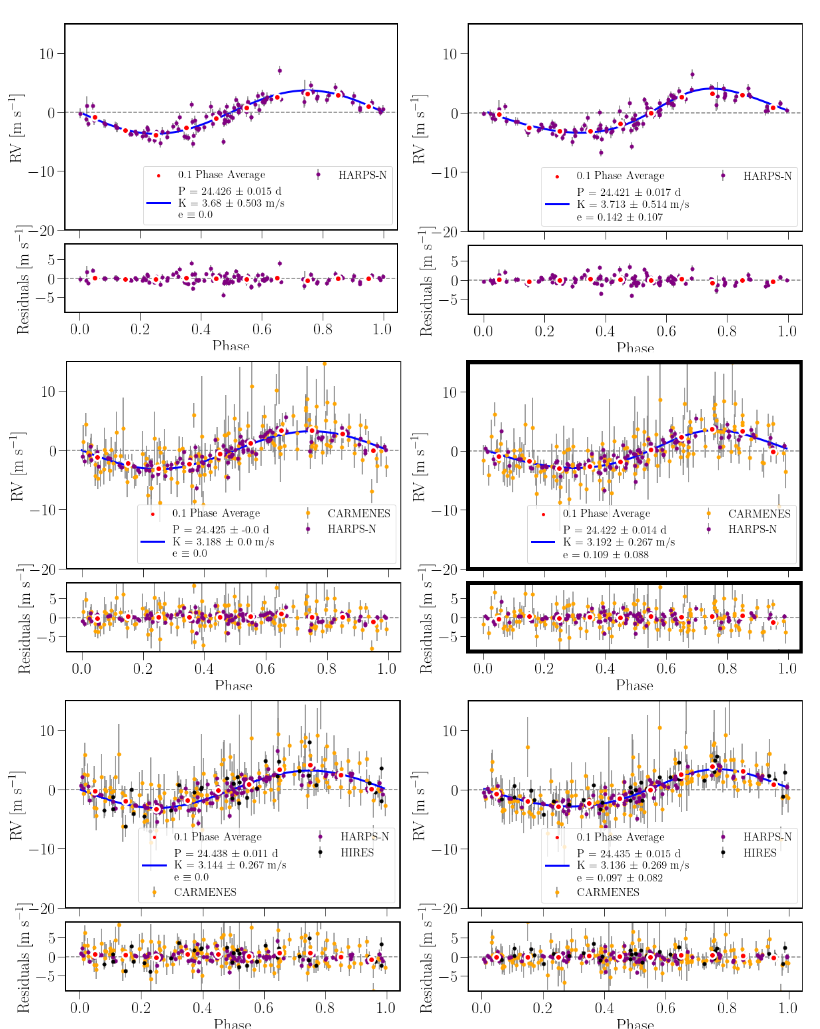}
\centering
\caption{Phase-folded, best-fit Keplerian orbital model, after the GPs and binary star trend have been removed, as well as their residuals. The top row shows the best-fit model to the HARPS-N data alone, the middle row shows the combined HARPS-N + CARMENES dataset, and the bottom row shows the HARPS-N + CARMENES + HIRES dataset. The left hand column shows the circular fits to these datasets and the right hand column shows the corresponding eccentric fits. The blue line is the best-fit model with the orbital parameters printed in the legend and listed in Table \ref{tab:RV_fits}. Red circles are the same velocities binned in 0.1 units of orbital phase. The middle, right hand panel, marked with a bold border, shows our final, adopted parameters for the planet candidate.}
\label{fig:phasefolded_all}
\end{figure*}

{Using \texttt{PyORBIT} \citep{malavolta_pyorbit_2016}, we fit the HARPS-N RVs of HD79211 with the priors described in the appendix (Section \ref{sect:RV_priors}).} We performed similar fits for HD79210 and were not able to find any planet signals. We report the results of six fits for HD79211: fitting just the HARPS-N radial velocities, fitting the combined HARPS-N + CARMENES RVs, and fitting the combined HARPS-N + CARMENES + HIRES RVs, with both a circular and an eccentric orbit and with GPs to model stellar activity. The priors and results of our fits to the RVs from HD79211 can be found in Table \ref{tab:RV_fits} and the phase-folded best-fit models are shown in Figure \ref{fig:phasefolded_all}. The results of all of these runs are consistent within their quoted uncertainties

\begin{figure}
\centering
\includegraphics[width=0.45\textwidth]{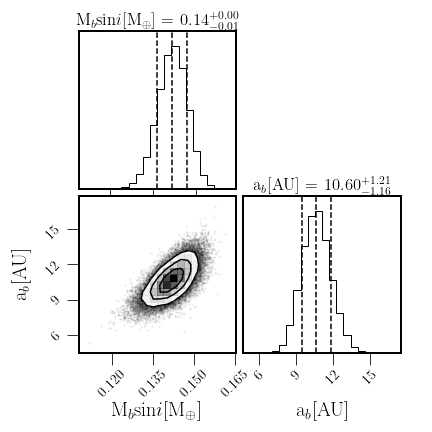}
\centering
\caption{Posterior distributions of the combined HARPS-N and CARMENES dataset for HD79211b's parameters derived from fitting an eccentric orbit.}

\smallskip
\smallskip
\smallskip
\smallskip
\smallskip
\smallskip

\label{fig:combined_ecc_corner_derived_params}
\end{figure}

We adopt the results of fitting the HARPS-N + CARMENES dataset with an eccentric orbit as our final results for the planet candidate around HD79211 (full posteriors shown in Figure \ref{fig:combined_ecc_corner}). This is the model favored by an AIC model comparison. Although the results of fitting the combination of all three datasets are consistent with those of fitting the HARPS-N + CARMENES dataset, an aliasing effect caused by a peak in the window function which is unique to the HARPS-N + CARMENES + HIRES dataset causes the posterior of the planet period to be double-peaked, at 24.44 and 24.38 days. Ultimately, we identify a signal with a period of $24.422\pm0.014$ days and attribute it to a planet candidate orbiting HD79211, with an $M$ sin $i$ of $10.6 \pm 1.2 M_\oplus$ and $a=0.142\pm0.005$ au (Figure \ref{fig:combined_ecc_corner_derived_params}).

\begin{figure}
\centering
\includegraphics[width=0.45\textwidth]{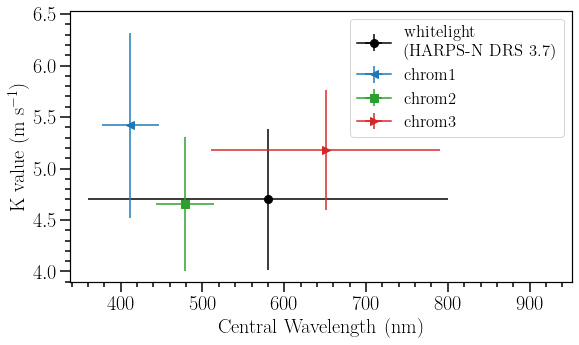}
\centering
\caption{We fit the chromatic radial velocities using \texttt{RadVel}, fixing all of the parameters and hyperparameters except for $K_b$ to the best-fit values found from fitting the white-light HARPS-N DRS RVs (the white light counterpart to the chromatic RVs), not the HARPS-TERRA RVs that we use in the rest of this work. Here, we plot the resulting $K_b$ versus the wavelength band used to calculate those radial velocities. The white light RV result from fitting the HARPS-N DRS RVs is noted by the black point. All of these values agree within their errorbars, and they do not show the trend of increasing semi-amplitude with decreasing wavelength that we would expect of a signal caused by stellar activity.}

\smallskip
\smallskip

\label{fig:chromatic_K}
\end{figure}

We also fit the chromatic RVs (described in section \ref{sect:HN_spec}) using \texttt{RadVel}, to see if the semi-amplitude of the signal in HD79211 varies with the wavelength of light used to calculate the radial velocities. Fixing the parameters and hyperparameters except for $K_b$ (i.e. $P_b$, $e_b$, $T_{conj}$, $Hamp$, $Pdec$, $Prot$, $Oamp$, $\sigma_{HN}$, c1, c2, x zero) to the best-fit values found from fitting the white-light HARPS-N DRS RVs, we fit the chromatic light curves and compare their semi-amplitudes. Note that we used the HARPS-N RVs as calculated by the DRS, not by HARPS-TERRA, to find these parameters. These results are shown in Figure \ref{fig:chromatic_K}. We found that the semi-amplitudes were consistent, within errorbars, across the chromatic RV sets and in comparison to the white light semi-amplitude. Although we caution against over interpreting these results, we could expect stellar activity to cause stronger signals in bluer light and smaller signals in redder light, whereas we expect a planet signal to be consistent across colors.

\begin{table*}
    \caption{RV fits for three sets of HD79211 RVs using \texttt{PyORBIT}. We fit HARPS-N RVs alone, HARPS-N + CARMENES RVs and HARPS-N + CARMENES + HIRES RVs with both eccentric and circular orbits. We adopt the results of fitting the HARPS-N + CARMENES RVs with an eccentric orbit as our final results. U in the prior column denotes a Uniform prior and G denotes a Gaussian prior.}
    \centering
    \begin{tabular}{llcccccccc}
            \noalign{\smallskip}
             &  &  & & &  & HARPS-N + & HARPS-N +  \\
             &  & HARPS-N  & HARPS-N  & HARPS-N +  & HARPS-N +  & CARMENES + &  CARMENES +  \\
             Parameter&\multicolumn{1}{c}{Prior}& Only &Only &CARMENES &CARMENES* & HIRES & HIRES \\
             & & $e=0$ & $e<0.3$ &$e=0$ &$e<0.3$ &$e=0$ &$e<0.3$ \\

            \noalign{\smallskip}
            \hline
            \noalign{\smallskip}
            
            Planet Orbit \\
            \noalign{\smallskip}
           
            $P_{b}$ (days) & U: 10-50 & $24.426_{-0.016}^{+0.015}$
            & $24.421^{+0.016}_{-0.017}$ & $24.425_{-0.014}^{+0.013}$ & $24.422\pm0.014$* & $24.438_{-0.012}^{+0.009}$ & $24.435_{-0.019}^{+0.011}$
            \\
            
            $K_{b}$ (m/s) & U: 0.01-10 & $3.68\pm0.50$
            & $3.71^{0.52}_{-0.51}$ & $3.19\pm0.27$ & $3.19\pm0.27$* & $3.14\pm0.27$ & $3.14\pm0.27$
            \\
            
            $e_{b}$ & Varies & $\equiv0.0$
            & $0.142^{+0.113}_{-0.100}$ & $\equiv0.0$ & $0.109 _{-0.075 }^{+0.100 }$* & $\equiv0.0$ & $0.097_{-0.067}^{+0.096}$
            \\
            
            $T\rm{conj}_{b}$ (JD-2.4e6) & G: $57517.06\pm6.15$ & $57522.08_{-0.54}^{+0.52}$
            & $57521.72^{ +1.02}_{-1.38} $ & $57521.99_{-0.36}^{+0.35}$ & $57521.43_{-0.90}^{+0.72}$* & $57521.81\pm0.35$ & $57521.34_{-0.83}^{+0.66}$
            \\

            \noalign{\smallskip}
            \hline
            \noalign{\smallskip}
            
            Derived \\
            \noalign{\smallskip}

            $a_b$ (au) & \multicolumn{1}{c}- & $0.142\pm0.005$
            & $0.142\pm0.005$ & $0.142\pm0.005$ & $0.142\pm0.005$* & $ 0.142\pm0.005 $ & $0.142\pm0.005$
            \\
            
            $M_b$ sin $i (M_\oplus)$ & \multicolumn{1}{c}- & $12.3_{-1.9}^{+2.0}$
            & $12.3^{+2.0}_{-1.8}$ & $10.7\pm1.2$ & $10.6\pm1.2$* & $10.6\pm1.2$ & $ 10.4\pm1.2$
            \\

            \noalign{\smallskip}
            \hline
            \noalign{\smallskip}
            
            Gaussian Process \\
            \noalign{\smallskip}
            
            Hamp HN (m/s) & \multicolumn{1}{c}- & $4.00_{-0.65}^{+0.82}$
            & $4.03^{+0.85}_{-0.68}$ & $3.79_{-0.62}^{+0.77}$ & $3.74 _{-0.62 }^{+0.76}$* & $3.70_{-0.61}^{+0.75}$ & $3.67_{-0.61}^{+0.74} $
            \\
            
            Hamp CAR (m/s) & \multicolumn{1}{c}- & \multicolumn{1}{c}-
            & \multicolumn{1}{c}- & $3.71_{-0.70}^{+0.94}$ & $3.75_{-0.71}^{+0.97}$* & $3.70_{-0.70}^{+0.95}$ & $ 3.72_{-0.72}^{+0.96}$
            \\
            
            Hamp HIRES (m/s) & \multicolumn{1}{c}- & \multicolumn{1}{c}-
            & \multicolumn{1}{c}- & \multicolumn{1}{c}- & \multicolumn{1}{c}- & $5.16_{-2.36 }^{+1.57}$ & $5.23_{-2.44}^{+1.54}$
            \\
            
            Pdec (days) & U: 15-1000 & $180_{-68}^{+69}$
            & $197_{-76}^{+89}$ & $200_{-40}^{+47}$ & $203_{-43}^{+49}$* & $209_{-43 }^{+49 }$ & $213_{-45}^{+51}$
            \\
            
            Prot (days) & U: 16.419-18.623 &  $16.70_{-0.09}^{+0.22}$
            & $16.75_{-0.12}^{+0.29}$ & $16.60\pm0.05$ & $16.58 _{-0.06}^{+0.05}* $ & $16.59\pm0.05$ & $16.59\pm0.05$
            \\
            
            Oamp & G: 0.5$\pm$0.05 & $0.48\pm0.05$
            & $0.48\pm0.05$ & $0.46\pm0.05$ & $0.46\pm0.05$* & $0.46\pm0.05$ & $0.46\pm0.05$
            \\
            
            \noalign{\smallskip}
            \hline
            \noalign{\smallskip}
            
            Binary Trend \\
            \noalign{\smallskip}            

            c1: & \multicolumn{1}{c}- & $-0.01\pm0.001$
            & $-0.01\pm0.001$ & $-0.009\pm0.001$ & $-0.01\pm0.001$* & $-0.002\pm0.001$ & $-0.009\pm0.001$
            \\
            
            c2 (e-07): & \multicolumn{1}{c}- & $-1.86_{-1.17}^{+1.14}$
            & $-1.94_{-1.16}^{+1.13}$ & $-1.04_{-1.25}^{+1.23}$ & $-1.06_{-1.26}^{+1.21}$* & $-.46_{-1.40}^{+1.33}$ & $-.54_{-1.45}^{+1.35}$
            \\
            
            x zero (JD-2.4e6): & \multicolumn{1}{c}- & 57773.91
            & 57871.94 & 57871.94 & 57871.94* & 57773.91 & 57773.91
            \\

            \noalign{\smallskip}
            \hline
            \noalign{\smallskip}
            
            Offset and Jitter \\
            \noalign{\smallskip} 
            
            Off HN (m/s): & \multicolumn{1}{c}- & $0.63_{-1.12}^{+1.25}$
            & $-0.38_{-1.14}^{+1.26}$ & $-0.59_{-1.09}^{+1.13}$ & $-0.61_{-1.07}^{+1.12}$* & $ 0.22_{-1.10}^{+1.14 }$ & $0.23_{-1.09}^{+1.12}$
            \\
            
            Off CAR (m/s): & \multicolumn{1}{c}- & $0.81\pm0.37$
            & $-0.20\pm0.36$ & $-0.19_{-1.50}^{+1.51}$ & $-0.17_{-1.52}^{+1.54}$* & $ 0.72_{-1.50}^{+1.52}$ & $0.74_{-1.52}^{+ 1.55} $
            \\
            
            Off HIRES (m/s): & \multicolumn{1}{c}- & $-46.49\pm2.59$
            & $-47.59_{-2.67}^{+2.64}$ & $-45.93_{-2.90}^{+2.81}$ & $ -46.06_{-2.82}^{+2.86}$* & $-45.42_{-3.21}^{+3.19}$ & $-45.43_{-3.23}^{+3.17}$
            \\
            
            Jit HN (m/s): & \multicolumn{1}{c}- & $1.47_{-0.19}^{+0.20}$
            & $1.49_{-0.19}^{+0.20}$ & $1.51_{-0.18}^{+0.20}$ & $1.53_{-0.19}^{+0.20}$* & $1.54_{-0.18}^{+0.20}$ & $1.55_{-0.18}^{+0.20}$
            \\
            
            Jit CAR (m/s): & \multicolumn{1}{c}- & $3.88_{-0.28}^{+0.30}$
            & $3.88_{-0.28}^{+0.30}$ & $1.66\pm-0.28$ & $1.64\pm0.28$* & $1.67\pm0.28$ & $ 1.65_{-0.27}^{+0.28}$
            \\
            
            Jit HIRES (m/s): & \multicolumn{1}{c}- & $5.60_{-0.73}^{+0.91}$
            & $5.58_{-0.72}^{+0.89}$ & $5.63_{-0.73}^{+0.89}$ & $5.63_{-0.73}^{+0.90}$* & $3.02_{-1.88}^{+2.13} $ &  $ 2.99_{-1.91}^{+2.16}$
            \\
            
    \end{tabular}
    \label{tab:RV_fits}
        \footnotesize{\\
        * Adopted fit values}
        
\smallskip
\smallskip
\smallskip

\end{table*}


\section{Results and Discussion}\label{sect:discussion}

We present the results of analyzing HARPS-N, CARMENES and HIRES radial velocities from HD79210 and HD79211. We ultimately conclude that there likely is a planet orbiting HD79211, with a period of $24.422\pm0.014$ days, semi-amplitude of $3.19\pm0.27$ m/s, $M$ sin $i = 10.6 \pm 1.2 M_\oplus$, and $a = 0.142\pm0.005$ au, and find no evidence of any planet candidates around HD79210.

There are discrepancies between the periodograms of HD79211's radial velocities from HARPS-N alone (Figure \ref{fig:pgram_RVonly_11}) and CARMENES alone \citep{gonzalez-alvarez_carmenes_2020}. Periodograms of the CARMENES RVs show three peaks significant above the 0.1\% FAP level at 8.3, 24.4 and 16.6 days, explained by \cite{gonzalez-alvarez_carmenes_2020} as a harmonic of the stellar rotation period, the planet's orbital period and the stellar rotation period, respectively. The periodograms of the HARPS-N RVs, however, show two significant peaks above the 0.01\% FAP level at 24.41 and 24.76 days. We suspect that these discrepancies are caused by the fluctuation in stellar activity between seasons. Due to the low time-density of HARPS-N data, however, we are not able to isolate the HARPS-N data from the seasons where the two datasets overlap in order to perform a direct comparison.

All of our RV fits (Table \ref{tab:RV_fits}) found a periodic signal at $\approx 24.4$ days, with all parameters agreeing within their errorbars. Ultimately, this work supports the detection of the 24.4 day planet candidate around HD79211 originally published by \cite{gonzalez-alvarez_carmenes_2020}, and finds no additional candidates around either HD79211 or its binary companion HD79210. As discussed in that work, this planet candidate is one of the lowest mass planets discovered orbiting one member of a stellar binary with a semi-major axis below 400 au. It likely lies on the inner edge of its star's habitable zone, but it is likely non-transiting, with only a $5^{+2}_{-1}\%$ chance that it is transiting \citep{gonzalez-alvarez_carmenes_2020}. Even so, with these two stars' borderline-wide separation of a=130 au, this system can serve as a valuable datapoint in future studies of exoplanet formation and occurance around stellar binaries.

\section*{Acknowledgements}
This material is based upon work supported by the National Science Foundation Graduate Research Fellowship under Grant No. DGE1745303. The HARPS-N project was funded by the Prodex Program of the Swiss Space Office (SSO), the Harvard- University Origin of Life Initiative (HUOLI), the Scottish Universities Physics Alliance (SUPA), the University of Geneva, the Smithsonian Astrophysical Observatory (SAO), the Italian National Astrophysical Institute (INAF), University of St. Andrews, Queen’s University Belfast, and University
of Edinburgh. Parts of this work have been supported by the National Aeronautics and Space Administration under grant No. NNX17AB59G, issued through the Exoplanets Research Program. Parts of this work have been supported by the Brinson Foundation. R.D.H. is funded by the UK Science and Technology Facilities Council (STFC)'s Ernest Rutherford Fellowship (grant number ST/V004735/1). T.G.W and A.C.C acknowledge support from STFC consolidated grant numbers ST/R000824/1 and ST/V000861/1, and
UKSA grant ST/R003203/1.
This work has made use of data from the European Space Agency (ESA) mission
{\it Gaia} (\url{https://www.cosmos.esa.int/gaia}), processed by the {\it Gaia}
Data Processing and Analysis Consortium (DPAC,
\url{https://www.cosmos.esa.int/web/gaia/dpac/consortium}). Funding for the DPAC
has been provided by national institutions, in particular the institutions
participating in the {\it Gaia} Multilateral Agreement.


\appendix

\section{Selection of Priors for Radial Velocity Fits}\label{sect:RV_priors}

When fitting for a planet candidate around HD79211, we used the time of conjunction for the 24.4 day planet candidate published in \cite{gonzalez-alvarez_carmenes_2020} as our Gaussian prior (2457517.06 $\pm$ 6.15 days). We ran both circular-orbit fits and eccentric fits. We started with a wide prior on eccentricity ($e<0.8$), but found that it favored highly eccentric and physically unlikely orbits, resulting from fitting a gap in data with the peak RV. We then narrowed to $e<0.3$. We kept the prior on our semi-amplitude ($K_b$) wide throughout our fits, with a Jeffery's prior from 0.01-10 m/s.

We initially set a wide, uniform prior on orbital period (3-50 days), which resulted in a double-peaked posterior, with one peak tending toward the lower end of this prior range and another at around $\approx 25$ days. After constraining the orbital period to the lower range, we found that the lower peak resulted from the code trying to fit a one day periodic signal which originates in the window function. We then opted to increase the lower bound of our orbital period prior (uniform prior from $10$-$50$ days), preventing the MCMC from fitting that one-day periodicity.

In our fits using Gaussian processes to model stellar activity, we used a quasi-periodic kernel with four hyperparameters: variability amplitude (Hamp), non-periodic characteristic length (associated with the spot decay timescale, Pdec), variability period (associated with the stellar rotation period, Prot) and periodic characteristic length (associated with the number of spots/spot regions on the surface of the star, Oamp).

We set a wide uniform prior on Pdec from 15.0-1000.0 days. The lower limit is associated with the approximate lower limit on the estimated rotation period of the star, since if the stellar magnetically active regions are evolving more quickly than the star is rotating, we would not see a periodic signal at all. The upper limit is many times the rotation period of the star, to allow for a fit to stable, long-lived magnetically active regions on the star.

We then set a uniform prior on Prot from 16.419-18.623 days, restricting this value to the one-standard deviation range around the peaks in the periodograms of the light curves for HD79211 (see Section \ref{sect:LC_analysis}). We set a Gaussian prior of $0.5 \pm 0.05$ on Oamp \citep{jeffers_analytical_2009,haywood_accurate_2018}. This accounts for the fact that the photometric and RV variability effect of even highly complex spot distributions on a star will average out to those caused by two or three large spots on the visible surface of the star.

When fitting the chromatic HARPS-N radial velocities, we aim to detect differences in the semi-amplitude of the periodic signal at the suspected planet's period. To do this, we decided to fix all of the free parameters - $T_{conj}$, $P_b$, Hamp, Pdec, Prot, Oamp, c1, c2, x zero - to the best-fit values returned from fitting the white-light HARPS-N DRS and CARMENES joint data set. Note that, for this step, we used the HARPS-N RVs as calculated by the DRS (which are the white light counterpart to the chromatic radial velocities), and not those calculated by HARPS-TERRA, to find these parameters. Although we do expect the chromatic nature of stellar activity to cause changes in the amplitude of the stellar activity signal (Hamp) between chromatic sets of RV data, we opted to fix this value in order to remove the degeneracy between the change in amplitude induced by stellar activity and by the planet candidate.

\section{Figures}

\setcounter{figure}{0}
\renewcommand{\thefigure}{A\arabic{figure}}

\begin{figure}
\centering
\includegraphics[width=0.48\textwidth]{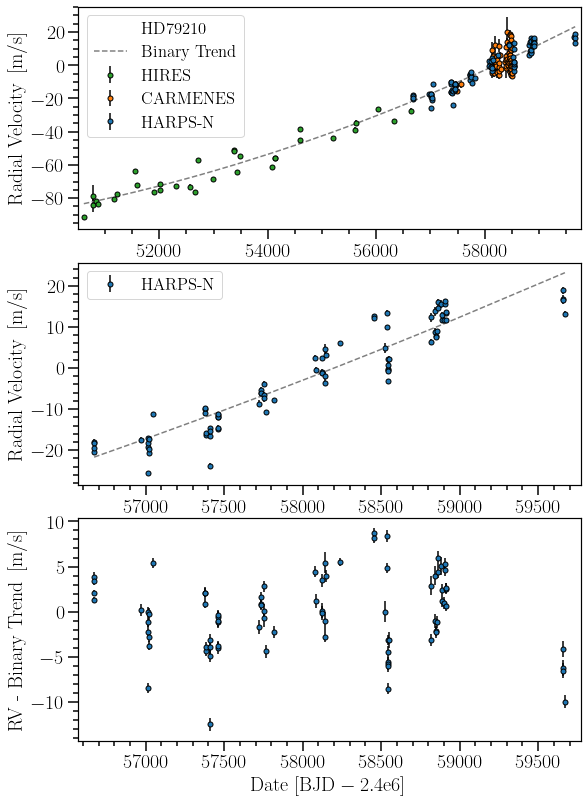}
\centering
\caption{{Radial velocities for HD79210 measured from CARMENES (orange, top panel only), HARPS-N (blue, all panels) and HIRES (green, top panel only) spectra. For most data points, the uncertainty in the RV is smaller than the size of the point. We mark the binary trend, fit as a second-order polynomial, with a dashed line in the first and third panels. The bottom panel shows the HARPS-N radial velocities with the binary trend subtracted. We use the HARPS-N RVs with the binary trend removed in our periodograms and when calculating correlation with stellar activity indicators analysis} (Figures \ref{fig:act_inds_10}, \ref{fig:pgram_RVonly_10}, \ref{fig:HN_RV_pgram_10}).}
\label{fig:HN_RVs_10}
\end{figure} 

\begin{figure}
\centering
\includegraphics[width=0.5\textwidth]{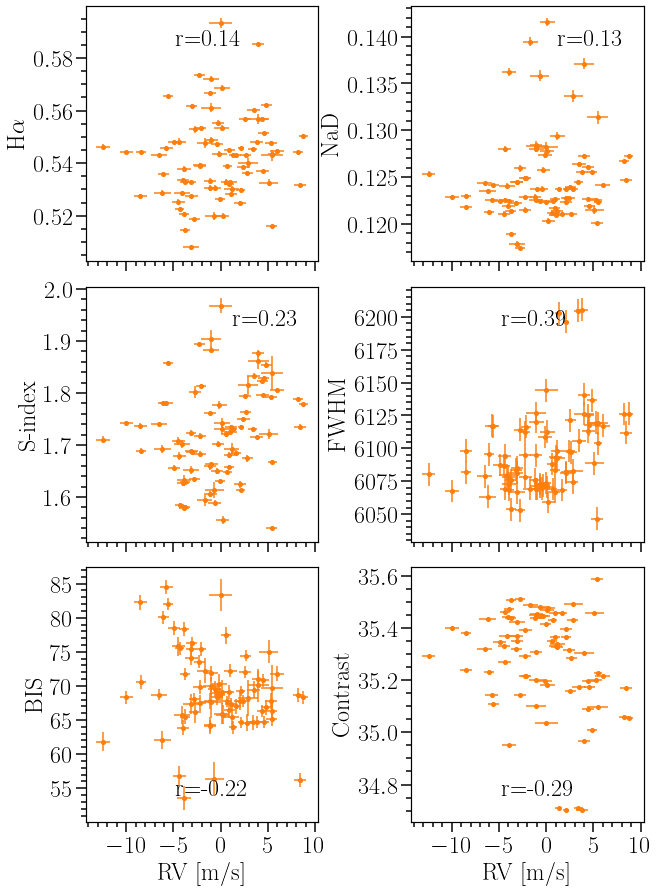}
\centering
\caption{Various HARPS-N stellar activity indicators versus HD79210's radial velocities after the stellar binary trend has been subtracted. The r-values are Pearson correlation coefficients. FWHM is moderately correlated ($r\geq 0.3$), with an r-value of 0.39.}
\label{fig:act_inds_10}
\end{figure} 

\begin{figure*}
    \centering
    \includegraphics[width=0.95\textwidth]{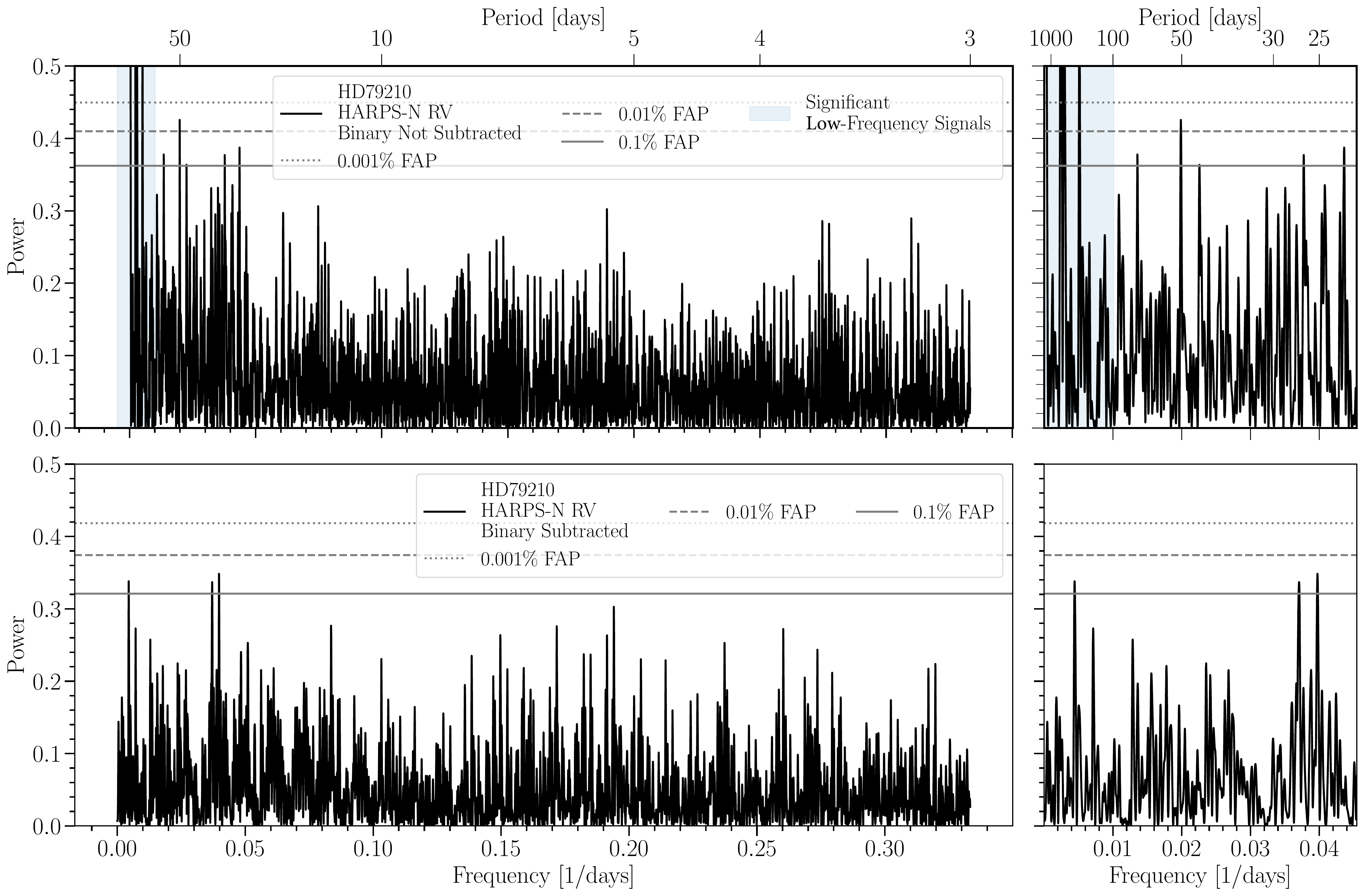}
    \caption{{Periodograms of HARPS-N radial velocities of HD79210, without the binary trend removed (top) and with the binary trend removed (bottom). Signals with frequencies from 0 to 0.33 $\text{ days}^{-1}$ are plotted. Before the binary trend is removed, there are many significant low frequency signals (blue shaded region), as we would expect from the long period binary. Once the binary trend is removed, there are no signals with less than 0.01\% FAP.}
    \smallskip
\smallskip
\smallskip}
    \label{fig:pgram_RVonly_10}
\end{figure*}

\begin{figure}
\centering
\includegraphics[width=0.45\textwidth]{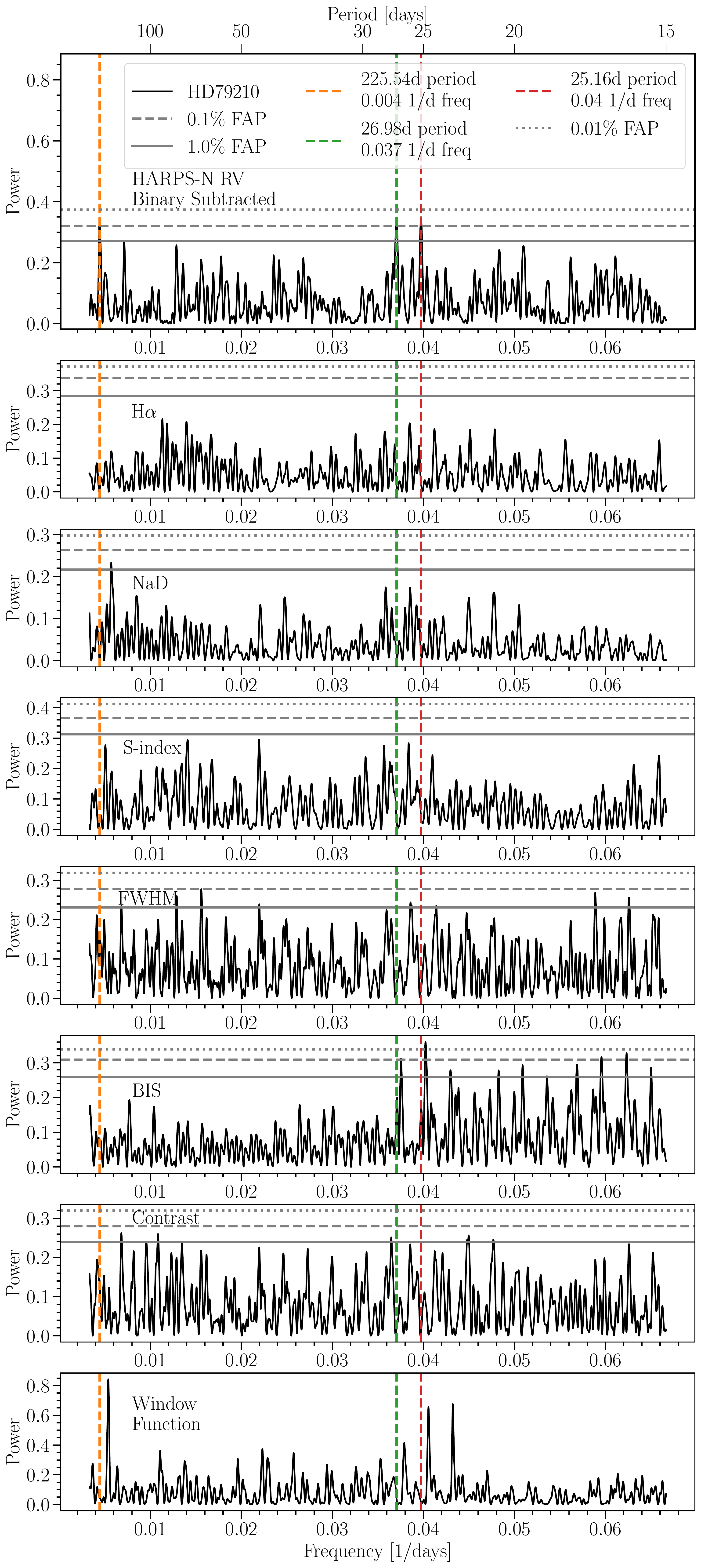}
\centering
\caption{Periodograms of HD79210's white light HARPS-N radial velocities with the binary trend removed, activity indicators, and the window function. The three most significant peaks ($<0.1\%$ FAP) are marked with dashed vertical lines. Note that there are no periodic signals in the RVs with FAP$<0.01\%$. The 25.16 day and 26.98 day signals are just offset from significant ($<0.01\%$ FAP and $<0.1\%$ FAP, respectively) peaks in the periodogram of BIS. We suspect that the RV signals correspond to these BIS signals, as BIS signals has been shown to be offset in time from RV variability \citep{dumusque_soap_2014,collier_cameron_three_2019}. We suspect that the 225.54 day signal is a harmonic of the 25.16 day variability signal.}
\label{fig:HN_RV_pgram_10}
\end{figure} 

\begin{figure*}
\centering
\includegraphics[width=\textwidth]{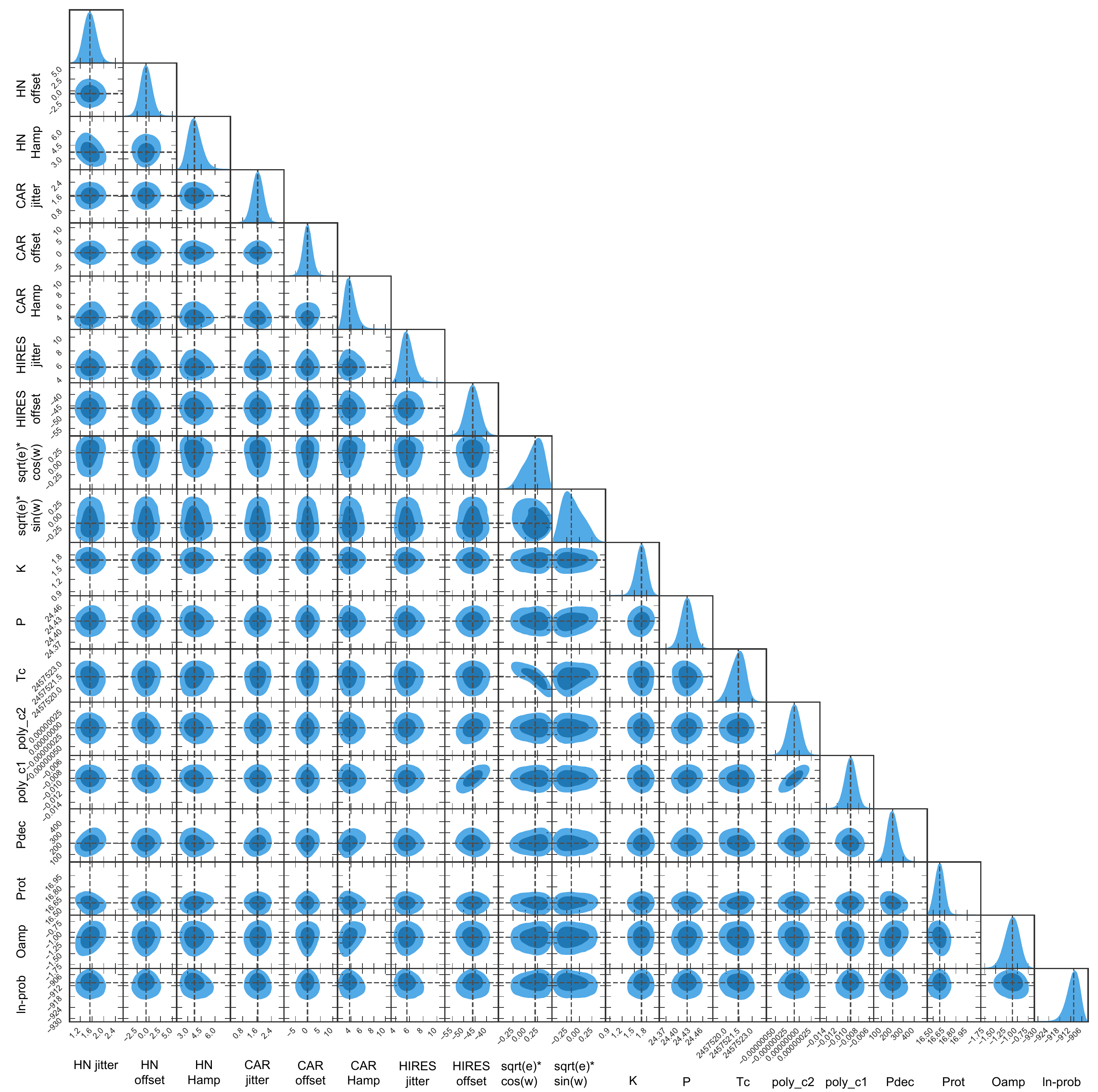}
\centering
\caption{Posterior distributions of the free parameters when fitting HD79211's combined HARPS-N and CARMENES RV dataset with an eccentric orbit and a quasic-periodic GP kernel.}


\label{fig:combined_ecc_corner}
\end{figure*}

\section{Tables}

\setcounter{table}{0}
\renewcommand{\thetable}{B\arabic{table}}

\begin{table*}
    \caption{All available magnitudes from survey photometry for HD79210 and HD79211. We also note the flags associated with these photometry, which denote the extent to which the two sources were blended in the original images.}
    \centering
\begin{tabular}{lllll}

\hline
\noalign{\smallskip}

\multicolumn{1}{c}{Survey}                               & \multicolumn{1}{c}{Band} & \multicolumn{1}{c}{HD79210} & \multicolumn{1}{c}{HD79211} & \multicolumn{1}{c}{Flags}   \\                         

\noalign{\smallskip}
\hline
\noalign{\smallskip}

The Hipparcos and Tycho Catalogues$^\text{a}$            & BTmag                    & 9.388 $\pm$ 0.03            &                             &\\
The Hipparcos and Tycho Catalogues$^\text{a}$            & VTmag                    & 7.791 $\pm$ 0.016           &                             &\\
The Hipparcos and Tycho Catalogues$^\text{a}$            & Hpmag                    & 7.7796 $\pm$ 0.0728         & 7.9664 $\pm$ 0.0824         &\\
The Hipparcos and Tycho Catalogues$^\text{a}$            & Vmag                     & 7.64                        & 7.7                         &\\
Gaia EDR3$^\text{b}$                     & Gmag                     & 6.976026                    & 7.054455                    &\\
Gaia EDR3$^\text{b}$                     & BPmag                    & 7.863408                    & 7.959406                    &\\
Gaia EDR3$^\text{b}$                     & RPmag                    & 6.046104                    & 6.112957                    &\\
GALEX-DR5$^\text{c}$ & FUV                      & 19.471 $\pm$ 0.166          & 19.467 $\pm$ 0.165          & 1\\
GALEX-DR5$^\text{c}$ & NUV                      & 16.317 $\pm$ 0.016          & 16.35 $\pm$ 0.016           & 1\\
The Tycho-2 Catalogue$^\text{d}$                        & Tycho-2 BT mag           & 9.412                       & 9.549                       &\\
The Tycho-2 Catalogue$^\text{d}$                        & Tycho-2 VT mag           & 7.789                       & 7.882                       &\\
2MASS All-Sky Catalog of Point Sources$^\text{e}$     & Jmag                     & 4.889 $\pm$ 0.037           & 4.779 $\pm$ 0.174           & 2,3 \\
2MASS All-Sky Catalog of Point Sources$^\text{e}$     & Hmag                     & 3.987 $\pm$ 0.188           & 4.043 $\pm$ 0.206           & 3\\
2MASS All-Sky Catalog of Point Sources$^\text{e}$     & Kmag                     & 3.988 $\pm$ 0.036           & 4.136 $\pm$ 0.020        "  & 2,3 \\
The USNO-B1.0 Catalog$^\text{f}$                      & B1mag                    & 9.28                        & 9.41                        &\\
The USNO-B1.0 Catalog$^\text{f}$                      & B2mag                    & 6.85                        & 6.92                        &\\
The USNO-B1.0 Catalog$^\text{f}$                      & R1mag                    & 8.35                        & 8.46                        &\\
The USNO-B1.0 Catalog$^\text{f}$                      & R2mag                    & 6.78                        & 6.85                        &\\
SDSS DR13$^\text{g}$                                & u                        & 13.529892 $\pm$ 0.0379      & 12.307867 $\pm$ 0.0102      & 4\\
SDSS DR13$^\text{g}$                                & g                        & 14.218927 $\pm$ 0.2757      & 8.6298 $\pm$ 2.83E-04       & 4\\
SDSS DR13$^\text{g}$                                & r                        & 14.097498 $\pm$ 1.0859      & 7.2558894 $\pm$ 1.33E-04    & 4\\
SDSS DR13$^\text{g}$                                & i                        & 13.7924185 $\pm$ 1.0859     & 6.7036 $\pm$ 1.68E-04       & 4                                                                                                                                        \\
SDSS DR13$^\text{g}$                                & z                        & 11.406796 $\pm$ 1.0869      & 6.7336 $\pm$ 6.48E-04       & 4         \\                              

\noalign{\smallskip}
\hline
\noalign{\smallskip}

    \end{tabular}
    \label{tab:survey_phot}
    \footnotesize{\\ Flags: $^1$ The object was originally blended with another one $^2$ This category includes detections where the goodness-of-fit quality of the profile-fit photometry was very poor $^3$ Diffraction spike confusion $^4$ unclean
    
    References: $^\text{a}$ \cite{esa_hipparcos_1997} $^\text{b}$ \cite{gaia_collaboration_vizier_2020} $^\text{c}$ \cite{bianchi_galex_2011} $^\text{d}$ \cite{hog_tycho-2_2000} $^\text{e}$ \cite{cutri_vizier_2003} $^\text{f}$ \cite{monet_usno-b_2003} $^\text{g}$ \cite{blanton_sloan_2017}}

\end{table*}


\clearpage



    \clearpage
\begin{longtable}{rrrrrrrrrrrrrrr}

\toprule
        Date &    RV &  eRV &   FWHM &  eFWHM &  BIS &  eBIS &  Contrast &  eContrast &  Smw &  eSmw &    Ha &   eHa &     Na &    eNa \\
        (jdb-2.45e6) & (m s$^{-1}$) & (m s$^{-1}$) \\
\midrule
6285.612891 &  20.3 &  0.8 & 6117.3 &    8.7 & 61.8 &   1.0 & 34.232862 &   0.000001 & 1.869 &  0.004 & 0.577 & 0.001 & 0.1229 & 0.0002 \\
6671.615209 &   8.7 &  0.3 & 6029.1 &    8.5 & 56.8 &   1.0 & 34.624995 &   0.000001 & 0.000 &  0.000 & 0.551 & 0.001 & 0.1257 & 0.0002 \\
6671.724181 &   8.7 &  0.6 & 6025.1 &    8.5 & 58.0 &   1.1 & 34.621989 &   0.000001 & 0.000 &  0.000 & 0.545 & 0.001 & 0.1241 & 0.0002 \\
6672.512178 &  10.8 &  0.3 & 6040.1 &    8.5 & 57.7 &   1.1 & 34.550935 &   0.000001 & 0.000 &  0.000 & 0.556 & 0.001 & 0.1244 & 0.0002 \\
6672.609691 &  12.4 &  0.4 & 6133.3 &    8.7 & 60.7 &   0.9 & 34.136966 &   0.000001 & 1.724 &  0.003 & 0.558 & 0.001 & 0.1243 & 0.0002 \\
7014.621385 &  -0.2 &  0.6 & 5984.5 &    8.5 & 66.2 &   1.1 & 35.042534 &   0.000001 & 1.498 &  0.005 & 0.534 & 0.001 & 0.1206 & 0.0002 \\
7014.732420 &  -1.8 &  0.6 & 5984.8 &    8.5 & 68.9 &   1.0 & 35.034022 &   0.000001 & 1.480 &  0.004 & 0.533 & 0.001 & 0.1207 & 0.0002 \\
7018.726754 &   5.8 &  0.5 & 5994.5 &    8.5 & 67.2 &   1.2 & 34.991576 &   0.000001 & 1.437 &  0.006 & 0.523 & 0.001 & 0.1214 & 0.0003 \\
7020.717072 &   2.8 &  0.6 & 5990.4 &    8.5 & 74.4 &   1.2 & 35.018739 &   0.000001 & 1.581 &  0.006 & 0.540 & 0.001 & 0.1199 & 0.0003 \\
7020.785177 &   4.9 &  0.5 & 5991.4 &    8.5 & 72.8 &   0.9 & 34.986452 &   0.000001 & 1.536 &  0.004 & 0.532 & 0.001 & 0.1167 & 0.0002 \\
7044.748325 &   6.0 &  0.5 & 6002.1 &    8.5 & 72.3 &   0.9 & 34.923610 &   0.000001 & 1.561 &  0.004 & 0.535 & 0.001 & 0.1200 & 0.0002 \\
7045.530040 &   3.3 &  0.5 & 5998.6 &    8.5 & 68.7 &   1.2 & 34.967085 &   0.000001 & 1.542 &  0.006 & 0.545 & 0.001 & 0.1217 & 0.0003 \\
7045.722216 &   3.8 &  0.6 & 5995.6 &    8.5 & 68.6 &   1.3 & 34.983797 &   0.000001 & 1.536 &  0.007 & 0.535 & 0.001 & 0.1219 & 0.0003 \\
7046.602067 &   4.5 &  0.5 & 5997.0 &    8.5 & 68.1 &   0.9 & 34.963493 &   0.000001 & 1.518 &  0.004 & 0.528 & 0.001 & 0.1206 & 0.0002 \\
7046.737819 &   4.5 &  0.4 & 5995.3 &    8.5 & 67.9 &   0.9 & 34.966771 &   0.000001 & 1.502 &  0.003 & 0.532 & 0.001 & 0.1197 & 0.0002 \\
7048.470341 &   6.0 &  0.6 & 5989.9 &    8.5 & 62.4 &   1.3 & 35.022517 &   0.000001 & 1.471 &  0.007 & 0.518 & 0.001 & 0.1206 & 0.0003 \\
7048.557574 &   6.4 &  0.5 & 5988.5 &    8.5 & 63.3 &   1.4 & 35.022603 &   0.000001 & 1.474 &  0.007 & 0.520 & 0.001 & 0.1211 & 0.0003 \\
7051.507932 &  14.6 &  0.5 & 6006.5 &    8.5 & 65.8 &   1.1 & 34.896503 &   0.000001 & 1.526 &  0.005 & 0.531 & 0.001 & 0.1218 & 0.0002 \\
7051.628868 &  14.2 &  0.5 & 6009.1 &    8.5 & 65.9 &   1.0 & 34.883580 &   0.000001 & 1.603 &  0.004 & 0.549 & 0.001 & 0.1227 & 0.0002 \\
7052.598236 &   9.2 &  0.6 & 6014.2 &    8.5 & 71.9 &   1.4 & 34.869929 &   0.000001 & 1.546 &  0.008 & 0.539 & 0.001 & 0.1220 & 0.0003 \\
7052.673872 &   9.3 &  1.0 & 6004.5 &    8.5 & 68.8 &   2.3 & 34.884387 &   0.000001 & 1.551 &  0.016 & 0.535 & 0.002 & 0.1228 & 0.0005 \\
7053.433842 &   7.3 &  0.7 & 6008.7 &    8.5 & 72.7 &   1.5 & 34.889942 &   0.000001 & 1.588 &  0.009 & 0.547 & 0.001 & 0.1213 & 0.0003 \\
7053.572232 &   7.9 &  0.7 & 6007.6 &    8.5 & 72.1 &   1.3 & 34.881916 &   0.000001 & 1.612 &  0.007 & 0.553 & 0.001 & 0.1211 & 0.0003 \\
7054.587638 &   9.7 &  0.5 & 6009.7 &    8.5 & 68.7 &   0.9 & 34.781720 &   0.000001 & 1.710 &  0.004 & 0.560 & 0.001 & 0.1235 & 0.0002 \\
7054.675635 &   9.5 &  0.5 & 6005.7 &    8.5 & 70.3 &   0.9 & 34.885117 &   0.000001 & 1.590 &  0.003 & 0.547 & 0.001 & 0.1214 & 0.0002 \\
7106.458292 &   5.8 &  0.5 & 6002.2 &    8.5 & 69.3 &   1.2 & 34.941747 &   0.000001 & 1.434 &  0.005 & 0.504 & 0.001 & 0.1176 & 0.0002 \\
7106.581258 &   5.1 &  0.5 & 6005.6 &    8.5 & 68.8 &   1.2 & 34.936934 &   0.000001 & 1.435 &  0.006 & 0.500 & 0.001 & 0.1165 & 0.0002 \\
7107.383100 &   3.7 &  0.5 & 5996.2 &    8.5 & 71.5 &   0.8 & 34.948208 &   0.000001 & 1.462 &  0.003 & 0.525 & 0.001 & 0.1202 & 0.0002 \\
7107.511736 &   3.0 &  0.5 & 5996.9 &    8.5 & 73.1 &   1.0 & 34.955592 &   0.000001 & 1.452 &  0.005 & 0.512 & 0.001 & 0.1180 & 0.0002 \\
7324.740802 &   2.0 &  0.7 & 6013.2 &    8.5 & 71.9 &   1.4 & 34.906742 &   0.000001 & 1.612 &  0.008 & 0.537 & 0.001 & 0.1240 & 0.0003 \\
7324.746300 &   1.5 &  0.6 & 6001.4 &    8.5 & 71.3 &   1.4 & 34.895632 &   0.000001 & 1.599 &  0.008 & 0.534 & 0.001 & 0.1231 & 0.0003 \\
7325.700001 &   2.5 &  0.5 & 6004.0 &    8.5 & 74.8 &   1.2 & 34.916968 &   0.000001 & 1.549 &  0.006 & 0.515 & 0.001 & 0.1208 & 0.0003 \\
7325.705464 &   2.4 &  0.6 & 6003.9 &    8.5 & 74.7 &   1.2 & 34.925492 &   0.000001 & 1.556 &  0.006 & 0.518 & 0.001 & 0.1209 & 0.0003 \\
7332.667583 &   0.6 &  0.6 & 5996.4 &    8.5 & 67.9 &   1.0 & 34.961330 &   0.000001 & 1.572 &  0.005 & 0.533 & 0.001 & 0.1225 & 0.0002 \\
7332.753053 &   1.7 &  0.6 & 5998.0 &    8.5 & 69.0 &   1.0 & 34.960201 &   0.000001 & 1.542 &  0.004 & 0.525 & 0.001 & 0.1225 & 0.0002 \\
7333.681902 &   1.3 &  0.4 & 6001.3 &    8.5 & 67.8 &   0.8 & 34.925854 &   0.000001 & 1.555 &  0.003 & 0.540 & 0.001 & 0.1245 & 0.0002 \\
7333.753366 &   2.0 &  0.5 & 6002.1 &    8.5 & 69.3 &   0.8 & 34.924959 &   0.000001 & 1.538 &  0.003 & 0.538 & 0.001 & 0.1248 & 0.0002 \\
7334.624478 &   0.6 &  0.7 & 6010.6 &    8.5 & 67.6 &   0.9 & 34.885663 &   0.000001 & 1.598 &  0.004 & 0.544 & 0.001 & 0.1236 & 0.0002 \\
7334.727032 &   2.3 &  0.7 & 6009.3 &    8.5 & 66.7 &   0.8 & 34.888516 &   0.000001 & 1.567 &  0.003 & 0.533 & 0.001 & 0.1224 & 0.0002 \\
7335.664700 &  -0.0 &  0.6 & 6008.8 &    8.5 & 69.4 &   1.6 & 34.925364 &   0.000001 & 1.553 &  0.010 & 0.517 & 0.001 & 0.1247 & 0.0004 \\
7335.751223 &  -7.1 &  0.6 & 6009.7 &    8.5 & 68.5 &   1.0 & 34.878111 &   0.000001 & 1.536 &  0.004 & 0.526 & 0.001 & 0.1235 & 0.0002 \\
7379.570185 &  -1.7 &  0.6 & 5993.8 &    8.5 & 71.2 &   1.1 & 34.968125 &   0.000001 & 1.494 &  0.005 & 0.506 & 0.001 & 0.1207 & 0.0002 \\
7379.667493 &  -1.5 &  0.4 & 5990.6 &    8.5 & 69.7 &   0.8 & 34.966462 &   0.000001 & 1.494 &  0.003 & 0.506 & 0.001 & 0.1208 & 0.0002 \\
7380.564210 &   0.1 &  0.6 & 5992.2 &    8.5 & 70.4 &   1.3 & 34.972041 &   0.000001 & 1.556 &  0.007 & 0.517 & 0.001 & 0.1217 & 0.0003 \\
7380.689505 &   1.0 &  0.5 & 5992.5 &    8.5 & 67.9 &   0.9 & 34.957374 &   0.000001 & 1.545 &  0.004 & 0.522 & 0.001 & 0.1213 & 0.0002 \\
7381.620483 &   0.5 &  0.4 & 5994.3 &    8.5 & 69.8 &   1.2 & 34.972357 &   0.000001 & 1.522 &  0.006 & 0.515 & 0.001 & 0.1218 & 0.0003 \\
7381.708704 &   1.2 &  0.3 & 5994.9 &    8.5 & 65.6 &   0.9 & 34.942763 &   0.000001 & 1.533 &  0.004 & 0.521 & 0.001 & 0.1225 & 0.0002 \\
7384.566887 &   1.8 &  0.6 & 6007.8 &    8.5 & 68.6 &   1.0 & 34.896642 &   0.000001 & 1.555 &  0.005 & 0.532 & 0.001 & 0.1234 & 0.0002 \\
7384.665722 &  -0.7 &  0.5 & 6005.1 &    8.5 & 70.2 &   1.1 & 34.899071 &   0.000001 & 1.552 &  0.005 & 0.531 & 0.001 & 0.1240 & 0.0002 \\
7385.535905 &  -1.7 &  0.7 & 6014.2 &    8.5 & 70.7 &   1.8 & 34.896648 &   0.000001 & 1.592 &  0.011 & 0.518 & 0.001 & 0.1226 & 0.0004 \\
7385.677403 &  -0.3 &  0.5 & 6009.3 &    8.5 & 68.4 &   1.2 & 34.863785 &   0.000001 & 1.618 &  0.006 & 0.527 & 0.001 & 0.1212 & 0.0002 \\
7407.475529 &   2.5 &  0.6 & 6036.3 &    8.5 & 71.0 &   0.9 & 34.716908 &   0.000001 & 1.686 &  0.004 & 0.567 & 0.001 & 0.1246 & 0.0002 \\
7407.717522 &   2.7 &  0.5 & 6037.7 &    8.5 & 72.8 &   0.9 & 34.710184 &   0.000001 & 1.642 &  0.004 & 0.556 & 0.001 & 0.1248 & 0.0002 \\
7408.512642 &  -0.8 &  0.5 & 6032.4 &    8.5 & 74.3 &   0.9 & 34.759374 &   0.000001 & 1.637 &  0.003 & 0.554 & 0.001 & 0.1238 & 0.0002 \\
7408.706636 &  -1.2 &  0.6 & 6029.6 &    8.5 & 80.3 &   1.2 & 34.772884 &   0.000001 & 1.668 &  0.006 & 0.550 & 0.001 & 0.1239 & 0.0003 \\
7409.461119 &  -1.0 &  0.6 & 6018.8 &    8.5 & 79.1 &   0.9 & 34.791592 &   0.000001 & 1.646 &  0.004 & 0.555 & 0.001 & 0.1237 & 0.0002 \\
7410.511099 &   1.6 &  0.9 & 6011.7 &    8.5 & 79.3 &   1.6 & 34.697860 &   0.000001 & 1.607 &  0.009 & 0.510 & 0.001 & 0.1210 & 0.0004 \\
7411.517177 &  -5.0 &  0.6 & 5992.5 &    8.5 & 72.5 &   1.2 & 34.905516 &   0.000001 & 1.567 &  0.005 & 0.517 & 0.001 & 0.1209 & 0.0002 \\
7411.761853 &  -1.4 &  0.8 & 5978.0 &    8.5 & 73.1 &   2.3 & 34.518417 &   0.000001 & 1.507 &  0.015 & 0.513 & 0.001 & 0.1296 & 0.0005 \\
7412.515048 &   3.3 &  0.5 & 5992.8 &    8.5 & 71.4 &   0.9 & 34.934547 &   0.000001 & 1.608 &  0.004 & 0.520 & 0.001 & 0.1192 & 0.0002 \\
7412.678358 &   2.9 &  0.5 & 5990.6 &    8.5 & 71.7 &   1.1 & 34.956586 &   0.000001 & 1.585 &  0.005 & 0.510 & 0.001 & 0.1188 & 0.0002 \\
7459.508092 &   3.1 &  0.8 & 6027.6 &    8.5 & 67.0 &   2.0 & 34.808038 &   0.000001 & 1.622 &  0.012 & 0.544 & 0.002 & 0.1233 & 0.0005 \\
7459.612820 &   2.0 &  0.8 & 6025.7 &    8.5 & 73.0 &   1.8 & 34.792587 &   0.000001 & 1.644 &  0.011 & 0.544 & 0.001 & 0.1231 & 0.0004 \\
7460.614946 &   2.9 &  0.5 & 6013.5 &    8.5 & 72.7 &   1.0 & 34.819873 &   0.000001 & 1.554 &  0.005 & 0.541 & 0.001 & 0.1235 & 0.0002 \\
7461.556115 &   2.8 &  0.6 & 6010.7 &    8.5 & 74.2 &   1.2 & 34.852056 &   0.000001 & 1.599 &  0.006 & 0.539 & 0.001 & 0.1231 & 0.0003 \\
7461.559865 &   1.9 &  0.5 & 6008.2 &    8.5 & 72.1 &   1.2 & 34.855077 &   0.000001 & 1.617 &  0.006 & 0.545 & 0.001 & 0.1240 & 0.0003 \\
7461.563661 &   1.6 &  0.6 & 6011.8 &    8.5 & 73.1 &   1.3 & 34.834035 &   0.000001 & 1.617 &  0.006 & 0.544 & 0.001 & 0.1235 & 0.0003 \\
7462.570242 &   1.8 &  0.7 & 6005.5 &    8.5 & 72.6 &   1.3 & 34.894691 &   0.000001 & 1.575 &  0.007 & 0.545 & 0.001 & 0.1228 & 0.0003 \\
7462.576318 &   1.7 &  0.7 & 6009.3 &    8.5 & 71.9 &   1.3 & 34.900651 &   0.000001 & 1.576 &  0.007 & 0.542 & 0.001 & 0.1225 & 0.0003 \\
7463.478321 &   6.4 &  0.6 & 5999.6 &    8.5 & 68.9 &   1.2 & 34.918489 &   0.000001 & 1.537 &  0.005 & 0.534 & 0.001 & 0.1220 & 0.0003 \\
7463.483760 &   5.0 &  0.6 & 5998.8 &    8.5 & 67.2 &   1.2 & 34.933660 &   0.000001 & 1.539 &  0.006 & 0.531 & 0.001 & 0.1222 & 0.0003 \\
7463.592711 &   5.2 &  0.7 & 6000.1 &    8.5 & 67.6 &   1.5 & 34.920505 &   0.000001 & 1.519 &  0.008 & 0.521 & 0.001 & 0.1211 & 0.0004 \\
7463.598243 &   5.1 &  0.8 & 5998.8 &    8.5 & 70.1 &   1.8 & 34.950547 &   0.000001 & 1.519 &  0.010 & 0.520 & 0.001 & 0.1210 & 0.0004 \\
7752.639756 &   0.9 &  0.8 & 6023.6 &    8.5 & 65.8 &   1.7 & 34.820135 &   0.000001 & 1.635 &  0.010 & 0.530 & 0.001 & 0.1273 & 0.0004 \\
7753.564001 &   1.3 &  1.1 & 6024.4 &    8.5 & 69.0 &   2.8 & 34.816407 &   0.000001 & 1.624 &  0.020 & 0.536 & 0.002 & 0.1371 & 0.0006 \\
7753.786752 &   1.9 &  0.6 & 6028.4 &    8.5 & 61.5 &   1.4 & 34.754533 &   0.000001 & 1.652 &  0.008 & 0.536 & 0.001 & 0.1244 & 0.0003 \\
8075.656797 &  -1.4 &  0.5 & 6005.1 &    8.5 & 71.8 &   0.9 & 34.921503 &   0.000001 & 1.572 &  0.003 & 0.519 & 0.001 & 0.1208 & 0.0002 \\
8119.563946 &  -3.8 &  0.9 & 5995.9 &    8.5 & 62.6 &   1.6 & 34.942547 &   0.000001 & 1.467 &  0.008 & 0.515 & 0.001 & 0.1201 & 0.0003 \\
8120.610033 &  -5.8 &  0.7 & 6002.2 &    8.5 & 62.6 &   1.6 & 34.907409 &   0.000001 & 1.480 &  0.008 & 0.551 & 0.001 & 0.1361 & 0.0004 \\
8122.652629 &  -0.7 &  0.7 & 5998.8 &    8.5 & 65.9 &   1.3 & 34.960728 &   0.000001 & 1.507 &  0.005 & 0.519 & 0.001 & 0.1203 & 0.0003 \\
8143.542711 &  -5.1 &  0.7 & 5983.7 &    8.5 & 74.7 &   1.6 & 35.038470 &   0.000001 & 1.463 &  0.010 & 0.506 & 0.001 & 0.1191 & 0.0003 \\
8147.611456 &  -3.2 &  0.7 & 5976.1 &    8.5 & 66.9 &   1.1 & 35.055493 &   0.000001 & 1.390 &  0.006 & 0.489 & 0.001 & 0.1186 & 0.0002 \\
8477.691075 & -13.5 &  0.9 & 6055.6 &    8.6 & 78.6 &   1.5 & 34.556499 &   0.000001 & 1.775 &  0.007 & 0.555 & 0.001 & 0.1250 & 0.0003 \\
8477.757699 & -13.7 &  0.8 & 6064.4 &    8.6 & 89.5 &   2.0 & 34.565657 &   0.000001 & 1.818 &  0.011 & 0.570 & 0.001 & 0.1267 & 0.0004 \\
8526.411740 &  -7.9 &  0.7 & 6076.2 &    8.6 & 76.5 &   1.3 & 34.461290 &   0.000001 & 1.881 &  0.006 & 0.581 & 0.001 & 0.1277 & 0.0003 \\
8536.630177 &   0.1 &  0.7 & 6042.3 &    8.5 & 60.1 &   1.2 & 34.635809 &   0.000001 & 1.699 &  0.005 & 0.532 & 0.001 & 0.1236 & 0.0002 \\
8538.634332 &   4.9 &  0.7 & 6054.7 &    8.6 & 65.7 &   1.2 & 34.566339 &   0.000001 & 1.743 &  0.006 & 0.546 & 0.001 & 0.1256 & 0.0003 \\
8805.645794 &  -3.5 &  0.9 & 6029.3 &    8.5 & 74.1 &   1.4 & 34.783666 &   0.000001 & 1.630 &  0.006 & 0.516 & 0.001 & 0.1250 & 0.0003 \\
8805.741588 &  -2.1 &  0.7 & 6026.6 &    8.5 & 72.6 &   0.9 & 34.739513 &   0.000001 & 1.623 &  0.003 & 0.519 & 0.001 & 0.1246 & 0.0002 \\
8816.605286 &  -5.7 &  0.7 & 6010.7 &    8.5 & 73.4 &   1.2 & 34.858948 &   0.000001 & 1.587 &  0.005 & 0.516 & 0.001 & 0.1223 & 0.0002 \\
8816.723732 &  -6.6 &  0.9 & 6017.1 &    8.5 & 69.9 &   1.5 & 34.879021 &   0.000001 & 1.589 &  0.007 & 0.511 & 0.001 & 0.1229 & 0.0003 \\
8818.625772 &  -6.5 &  0.7 & 6016.5 &    8.5 & 69.0 &   0.9 & 34.815676 &   0.000001 & 1.643 &  0.003 & 0.531 & 0.001 & 0.1245 & 0.0002 \\
8818.749021 &  -6.4 &  0.7 & 6017.3 &    8.5 & 67.1 &   0.9 & 34.812491 &   0.000001 & 1.613 &  0.003 & 0.519 & 0.001 & 0.1223 & 0.0002 \\
8830.770218 &  -4.4 &  0.8 & 6031.1 &    8.5 & 75.9 &   2.1 & 34.753263 &   0.000001 & 1.643 &  0.011 & 0.521 & 0.001 & 0.1253 & 0.0005 \\
8832.775143 &   0.9 &  0.8 & 6037.4 &    8.5 & 86.0 &   2.7 & 34.791074 &   0.000001 & 1.598 &  0.015 & 0.525 & 0.002 & 0.1443 & 0.0007 \\
8841.672974 &  -2.7 &  1.5 & 6024.0 &    8.5 & 53.6 &   4.0 & 34.995432 &   0.000001 & 1.776 &  0.029 & 0.539 & 0.003 & 0.1596 & 0.0009 \\
8847.594680 & -11.1 &  0.7 & 6018.0 &    8.5 & 75.5 &   1.0 & 34.836482 &   0.000001 & 1.573 &  0.004 & 0.513 & 0.001 & 0.1223 & 0.0002 \\
8847.756271 &  -9.0 &  0.8 & 6019.6 &    8.5 & 75.7 &   1.0 & 34.838014 &   0.000001 & 1.612 &  0.004 & 0.518 & 0.001 & 0.1218 & 0.0002 \\
8854.646741 &  -3.0 &  0.8 & 6008.0 &    8.5 & 69.4 &   0.9 & 34.899571 &   0.000001 & 1.579 &  0.003 & 0.543 & 0.001 & 0.1257 & 0.0002 \\
8883.554321 &  -7.5 &  1.0 & 6008.8 &    8.5 & 74.6 &   2.0 & 34.980193 &   0.000001 & 1.619 &  0.011 & 0.526 & 0.001 & 0.1214 & 0.0005 \\
8883.640535 &  -6.1 &  2.3 & 6010.4 &    8.5 & 67.4 &   4.8 & 35.207201 &   0.000001 & 1.783 &  0.041 & 0.517 & 0.003 & 0.1243 & 0.0010 \\
8887.479981 &  -4.4 &  0.7 & 6005.1 &    8.5 & 71.7 &   0.8 & 34.894021 &   0.000001 & 1.523 &  0.003 & 0.497 & 0.000 & 0.1184 & 0.0001 \\
8887.585245 &  -3.4 &  0.7 & 6004.0 &    8.5 & 71.6 &   0.8 & 34.900044 &   0.000001 & 1.547 &  0.003 & 0.505 & 0.000 & 0.1198 & 0.0001 \\
8908.546480 &  -4.2 &  0.7 & 6019.9 &    8.5 & 68.5 &   1.0 & 34.828600 &   0.000001 & 1.690 &  0.004 & 0.537 & 0.001 & 0.1208 & 0.0002 \\
8908.666000 &  -5.1 &  0.6 & 6020.9 &    8.5 & 66.0 &   1.3 & 34.832169 &   0.000001 & 1.678 &  0.006 & 0.535 & 0.001 & 0.1193 & 0.0003 \\
8910.496814 &  -2.2 &  0.7 & 6020.3 &    8.5 & 69.4 &   0.9 & 34.833626 &   0.000001 & 1.625 &  0.004 & 0.526 & 0.001 & 0.1216 & 0.0002 \\
8915.453070 &  -6.7 &  0.6 & 6009.5 &    8.5 & 71.8 &   0.8 & 34.885467 &   0.000001 & 1.536 &  0.003 & 0.515 & 0.000 & 0.1189 & 0.0002 \\
8915.546282 &  -9.3 &  0.8 & 6011.3 &    8.5 & 68.6 &   1.4 & 34.935664 &   0.000001 & 1.523 &  0.007 & 0.508 & 0.001 & 0.1187 & 0.0003 \\
9655.557594 & -19.1 &  0.7 & 6012.9 &    8.5 & 69.6 &   1.2 & 34.793949 &   0.000001 & 1.679 &  0.006 & 0.549 & 0.001 & 0.1235 & 0.0003 \\
9656.551881 & -19.9 &  1.1 & 6013.2 &    8.5 & 70.1 &   1.5 & 34.813970 &   0.000001 & 1.612 &  0.008 & 0.524 & 0.001 & 0.1235 & 0.0003 \\
9658.456802 & -19.7 &  0.9 & 6014.8 &    8.5 & 72.4 &   1.2 & 34.825784 &   0.000001 & 1.584 &  0.006 & 0.521 & 0.001 & 0.1227 & 0.0002 \\
9659.579979 & -18.2 &  0.7 & 6005.5 &    8.5 & 68.6 &   0.9 & 34.840340 &   0.000001 & 1.605 &  0.004 & 0.540 & 0.001 & 0.1230 & 0.0002 \\
9661.603197 & -15.8 &  0.9 & 6002.3 &    8.5 & 64.1 &   1.8 & 34.905553 &   0.000001 & 1.594 &  0.012 & 0.523 & 0.001 & 0.1211 & 0.0004 \\
9662.554508 & -13.6 &  0.8 & 6003.4 &    8.5 & 71.5 &   1.4 & 34.893858 &   0.000001 & 1.594 &  0.008 & 0.518 & 0.001 & 0.1209 & 0.0003 \\
\bottomrule

\caption{HD79211 radial velocities and activity indicators. RVs as calculated using TERRA, and activity indicators as calculated by the HARPS-N DRS 3.7.}

\end{longtable}
    \clearpage


    \clearpage
\begin{longtable}{rrrrrrrrrrrrrrr}
\toprule
        Date &    RV &  eRV &   FWHM &  eFWHM &  BIS &  eBIS &  Contrast &  eContrast &   Smw &  eSmw &    Ha &   eHa &     Na &    eNa \\
        (jdb-2.45e6) & (m s$^{-1}$) & (m s$^{-1}$) \\        
\midrule
6671.602883 & -17.9 &  0.6 & 6205.3 &    8.8 & 64.7 &   0.9 & 34.702755 &   0.000001 & 1.715 & 0.004 & 0.548 & 0.001 & 0.1255 & 0.0002 \\
6671.712040 & -18.3 &  0.5 & 6204.8 &    8.8 & 64.6 &   1.1 & 34.709181 &   0.000001 & 1.731 & 0.005 & 0.545 & 0.001 & 0.1245 & 0.0002 \\
6672.500523 & -20.4 &  0.4 & 6202.7 &    8.8 & 64.0 &   0.9 & 34.709411 &   0.000001 & 1.730 & 0.004 & 0.543 & 0.001 & 0.1237 & 0.0002 \\
6672.597364 & -19.5 &  0.3 & 6196.4 &    8.8 & 64.6 &   0.9 & 34.701540 &   0.000001 & 1.735 & 0.003 & 0.546 & 0.001 & 0.1237 & 0.0002 \\
6968.679139 & -17.4 &  0.7 & 6059.3 &    8.6 & 65.8 &   1.3 & 35.474655 &   0.000001 & 1.556 & 0.007 & 0.520 & 0.001 & 0.1204 & 0.0003 \\
7014.608861 & -19.2 &  0.7 & 6078.1 &    8.6 & 69.9 &   1.0 & 35.392864 &   0.000001 & 1.683 & 0.005 & 0.539 & 0.001 & 0.1228 & 0.0002 \\
7014.721910 & -25.5 &  0.5 & 6081.7 &    8.6 & 70.6 &   1.1 & 35.379604 &   0.000001 & 1.689 & 0.005 & 0.544 & 0.001 & 0.1229 & 0.0002 \\
7016.648683 & -17.1 &  0.6 & 6073.7 &    8.6 & 68.7 &   0.9 & 35.415204 &   0.000001 & 1.630 & 0.004 & 0.527 & 0.001 & 0.1224 & 0.0002 \\
7016.744011 & -18.1 &  0.6 & 6076.2 &    8.6 & 64.2 &   1.3 & 35.436469 &   0.000001 & 1.659 & 0.007 & 0.533 & 0.001 & 0.1229 & 0.0003 \\
7018.694229 & -17.3 &  0.7 & 6071.9 &    8.6 & 68.9 &   1.2 & 35.443635 &   0.000001 & 1.701 & 0.006 & 0.555 & 0.001 & 0.1258 & 0.0003 \\
7020.676238 & -20.7 &  0.5 & 6053.7 &    8.6 & 71.7 &   0.9 & 35.505181 &   0.000001 & 1.581 & 0.004 & 0.515 & 0.001 & 0.1189 & 0.0002 \\
7020.773116 & -19.7 &  0.5 & 6052.8 &    8.6 & 68.1 &   0.9 & 35.509273 &   0.000001 & 1.634 & 0.004 & 0.519 & 0.001 & 0.1174 & 0.0002 \\
7044.618278 & -11.2 &  0.6 & 6046.5 &    8.6 & 65.1 &   0.9 & 35.588323 &   0.000001 & 1.540 & 0.004 & 0.516 & 0.001 & 0.1201 & 0.0002 \\
7379.547614 &  -9.9 &  0.6 & 6081.2 &    8.6 & 67.8 &   1.3 & 35.395780 &   0.000001 & 1.624 & 0.007 & 0.525 & 0.001 & 0.1223 & 0.0003 \\
7379.558946 & -11.0 &  0.5 & 6083.7 &    8.6 & 69.1 &   1.0 & 35.365167 &   0.000001 & 1.657 & 0.005 & 0.532 & 0.001 & 0.1226 & 0.0002 \\
7379.650825 &  -9.8 &  0.5 & 6082.0 &    8.6 & 69.3 &   0.8 & 35.364910 &   0.000001 & 1.613 & 0.003 & 0.530 & 0.001 & 0.1228 & 0.0002 \\
7381.632555 & -16.3 &  0.5 & 6069.3 &    8.6 & 75.5 &   1.0 & 35.459527 &   0.000001 & 1.584 & 0.005 & 0.523 & 0.001 & 0.1226 & 0.0002 \\
7381.719966 & -15.8 &  0.5 & 6068.5 &    8.6 & 78.3 &   1.0 & 35.473583 &   0.000001 & 1.579 & 0.005 & 0.521 & 0.001 & 0.1224 & 0.0002 \\
7407.463307 & -16.5 &  0.7 & 6087.3 &    8.6 & 78.5 &   1.0 & 35.347004 &   0.000001 & 1.657 & 0.005 & 0.548 & 0.001 & 0.1224 & 0.0002 \\
7407.705901 & -14.6 &  0.6 & 6085.3 &    8.6 & 76.3 &   0.9 & 35.348105 &   0.000001 & 1.687 & 0.004 & 0.562 & 0.001 & 0.1245 & 0.0002 \\
7411.506298 & -23.9 &  0.7 & 6080.3 &    8.6 & 61.9 &   1.4 & 35.292921 &   0.000001 & 1.710 & 0.007 & 0.546 & 0.001 & 0.1253 & 0.0003 \\
7411.628022 & -15.4 &  0.8 & 6073.3 &    8.6 & 53.6 &   1.8 & 34.952576 &   0.000001 & 1.632 & 0.010 & 0.538 & 0.001 & 0.1362 & 0.0004 \\
7458.514008 & -11.5 &  0.7 & 6072.9 &    8.6 & 69.6 &   1.1 & 35.445024 &   0.000001 & 1.589 & 0.005 & 0.530 & 0.001 & 0.1224 & 0.0003 \\
7458.519656 & -11.9 &  0.7 & 6069.2 &    8.6 & 67.8 &   1.2 & 35.447314 &   0.000001 & 1.605 & 0.006 & 0.531 & 0.001 & 0.1226 & 0.0003 \\
7458.594999 & -11.8 &  0.7 & 6069.1 &    8.6 & 71.8 &   1.1 & 35.452089 &   0.000001 & 1.663 & 0.005 & 0.549 & 0.001 & 0.1237 & 0.0003 \\
7458.600508 & -11.2 &  0.6 & 6069.7 &    8.6 & 68.3 &   1.1 & 35.444495 &   0.000001 & 1.651 & 0.006 & 0.547 & 0.001 & 0.1237 & 0.0003 \\
7462.478835 & -14.8 &  0.7 & 6078.5 &    8.6 & 63.9 &   1.1 & 35.442927 &   0.000001 & 1.626 & 0.005 & 0.534 & 0.001 & 0.1219 & 0.0002 \\
7462.484310 & -14.5 &  0.6 & 6075.8 &    8.6 & 65.5 &   1.1 & 35.439044 &   0.000001 & 1.628 & 0.005 & 0.533 & 0.001 & 0.1214 & 0.0002 \\
7723.698932 &  -8.7 &  0.8 & 6069.4 &    8.6 & 72.2 &   1.9 & 35.488947 &   0.000001 & 1.593 & 0.011 & 0.548 & 0.002 & 0.1394 & 0.0005 \\
7732.568598 &  -6.0 &  0.6 & 6066.9 &    8.6 & 66.4 &   1.1 & 35.454995 &   0.000001 & 1.681 & 0.005 & 0.545 & 0.001 & 0.1217 & 0.0002 \\
7732.575995 &  -5.2 &  0.5 & 6068.3 &    8.6 & 67.0 &   1.1 & 35.455369 &   0.000001 & 1.685 & 0.005 & 0.543 & 0.001 & 0.1211 & 0.0002 \\
7732.732312 &  -6.2 &  0.5 & 6069.2 &    8.6 & 67.7 &   0.8 & 35.427932 &   0.000001 & 1.648 & 0.003 & 0.539 & 0.000 & 0.1225 & 0.0001 \\
7752.629201 &  -6.5 &  0.8 & 6070.5 &    8.6 & 65.9 &   1.5 & 35.467998 &   0.000001 & 1.730 & 0.009 & 0.553 & 0.001 & 0.1278 & 0.0004 \\
7753.542357 &  -7.3 &  1.0 & 6071.2 &    8.6 & 56.4 &   2.4 & 35.480444 &   0.000001 & 1.613 & 0.016 & 0.520 & 0.002 & 0.1358 & 0.0006 \\
7753.774564 &  -3.8 &  0.6 & 6074.2 &    8.6 & 64.6 &   1.2 & 35.429545 &   0.000001 & 1.674 & 0.006 & 0.536 & 0.001 & 0.1237 & 0.0003 \\
7764.562994 & -10.8 &  0.7 & 6094.4 &    8.6 & 56.8 &   1.5 & 35.267229 &   0.000001 & 1.679 & 0.008 & 0.548 & 0.001 & 0.1280 & 0.0004 \\
7817.518062 &  -7.9 &  0.6 & 6094.8 &    8.6 & 67.5 &   1.2 & 35.290803 &   0.000001 & 1.717 & 0.006 & 0.539 & 0.001 & 0.1215 & 0.0003 \\
8075.643996 &   2.6 &  0.6 & 6125.3 &    8.7 & 66.3 &   0.9 & 35.088364 &   0.000001 & 1.823 & 0.003 & 0.557 & 0.001 & 0.1261 & 0.0002 \\
8086.621865 &  -0.5 &  0.8 & 6092.6 &    8.6 & 65.5 &   1.9 & 35.338213 &   0.000001 & 1.692 & 0.012 & 0.530 & 0.001 & 0.1294 & 0.0004 \\
8120.622082 &  -1.0 &  0.8 & 6112.7 &    8.6 & 69.1 &   1.8 & 35.180743 &   0.000001 & 1.742 & 0.010 & 0.569 & 0.001 & 0.1415 & 0.0004 \\
8121.702763 &  -1.3 &  0.7 & 6108.6 &    8.6 & 70.2 &   1.5 & 35.197757 &   0.000001 & 1.776 & 0.008 & 0.544 & 0.001 & 0.1273 & 0.0003 \\
8122.640152 &   2.3 &  0.7 & 6105.8 &    8.6 & 69.4 &   1.2 & 35.174412 &   0.000001 & 1.834 & 0.006 & 0.560 & 0.001 & 0.1263 & 0.0003 \\
8142.664448 &   4.6 &  1.2 & 6118.9 &    8.7 & 69.7 &   3.3 & 35.097965 &   0.000001 & 1.838 & 0.033 & 0.543 & 0.002 & 0.1314 & 0.0007 \\
8143.527815 &  -3.6 &  0.6 & 6114.1 &    8.6 & 66.2 &   1.7 & 35.142612 &   0.000001 & 1.802 & 0.012 & 0.553 & 0.001 & 0.1259 & 0.0004 \\
8143.646565 &  -1.9 &  1.1 & 6126.8 &    8.7 & 67.6 &   2.4 & 35.101413 &   0.000001 & 1.904 & 0.018 & 0.561 & 0.002 & 0.1283 & 0.0005 \\
8147.598157 &   3.2 &  0.7 & 6140.6 &    8.7 & 71.1 &   1.1 & 34.965033 &   0.000001 & 1.876 & 0.007 & 0.585 & 0.001 & 0.1273 & 0.0002 \\
8239.368652 &   6.1 &  0.4 & 6103.9 &    8.6 & 66.4 &   0.9 & 35.225081 &   0.000001 & 1.668 & 0.004 & 0.543 & 0.001 & 0.1224 & 0.0002 \\
8454.691049 &  12.6 &  0.5 & 6126.2 &    8.7 & 68.3 &   0.9 & 35.053999 &   0.000001 & 1.778 & 0.003 & 0.551 & 0.001 & 0.1273 & 0.0002 \\
8454.736712 &  12.1 &  0.6 & 6126.0 &    8.7 & 68.6 &   1.0 & 35.057840 &   0.000001 & 1.789 & 0.004 & 0.544 & 0.001 & 0.1267 & 0.0002 \\
8526.425177 &   5.0 &  1.2 & 6144.3 &    8.7 & 83.4 &   2.4 & 35.034133 &   0.000001 & 1.968 & 0.014 & 0.593 & 0.002 & 0.1282 & 0.0005 \\
8536.616161 &  13.5 &  0.6 & 6111.9 &    8.6 & 56.2 &   1.0 & 35.167891 &   0.000001 & 1.735 & 0.005 & 0.532 & 0.001 & 0.1247 & 0.0002 \\
8538.621890 &  10.0 &  0.6 & 6136.6 &    8.7 & 66.9 &   1.1 & 35.008712 &   0.000001 & 1.854 & 0.005 & 0.562 & 0.001 & 0.1256 & 0.0002 \\
8544.459923 &  -0.4 &  0.6 & 6116.8 &    8.7 & 82.0 &   0.9 & 35.109791 &   0.000001 & 1.858 & 0.003 & 0.565 & 0.001 & 0.1242 & 0.0002 \\
8544.595796 &  -0.5 &  0.7 & 6117.2 &    8.7 & 84.5 &   1.0 & 35.143497 &   0.000001 & 1.780 & 0.005 & 0.546 & 0.001 & 0.1226 & 0.0002 \\
8545.458728 &  -3.2 &  0.7 & 6098.3 &    8.6 & 82.3 &   1.1 & 35.239870 &   0.000001 & 1.737 & 0.005 & 0.528 & 0.001 & 0.1218 & 0.0002 \\
8545.572797 &  -0.8 &  0.6 & 6095.5 &    8.6 & 80.0 &   1.0 & 35.231555 &   0.000001 & 1.781 & 0.004 & 0.536 & 0.001 & 0.1213 & 0.0002 \\
8546.411967 &   0.8 &  0.7 & 6086.8 &    8.6 & 75.8 &   1.4 & 35.331340 &   0.000001 & 1.708 & 0.007 & 0.525 & 0.001 & 0.1210 & 0.0003 \\
8546.513687 &   2.1 &  0.5 & 6083.9 &    8.6 & 75.4 &   1.1 & 35.317667 &   0.000001 & 1.723 & 0.005 & 0.533 & 0.001 & 0.1223 & 0.0002 \\
8548.397384 &   2.2 &  0.9 & 6066.5 &    8.6 & 67.4 &   1.4 & 35.421875 &   0.000001 & 1.652 & 0.008 & 0.508 & 0.001 & 0.1179 & 0.0003 \\
8816.630520 &  12.3 &  1.0 & 6082.9 &    8.6 & 68.2 &   2.7 & 35.492177 &   0.000001 & 1.815 & 0.018 & 0.540 & 0.002 & 0.1337 & 0.0006 \\
8818.637903 &   6.4 &  0.7 & 6081.4 &    8.6 & 74.0 &   0.9 & 35.367111 &   0.000001 & 1.688 & 0.003 & 0.527 & 0.001 & 0.1245 & 0.0002 \\
8841.694745 &  13.8 &  1.1 & 6126.3 &    8.7 & 70.1 &   2.4 & 35.303118 &   0.000001 & 1.862 & 0.015 & 0.557 & 0.002 & 0.1371 & 0.0006 \\
8843.649122 &   8.9 &  0.8 & 6120.0 &    8.7 & 70.0 &   1.3 & 35.200034 &   0.000001 & 1.883 & 0.006 & 0.572 & 0.001 & 0.1280 & 0.0003 \\
8847.607122 &   7.8 &  0.5 & 6116.3 &    8.6 & 75.4 &   1.1 & 35.215659 &   0.000001 & 1.813 & 0.005 & 0.553 & 0.001 & 0.1249 & 0.0002 \\
8847.769535 &   7.7 &  0.6 & 6112.1 &    8.6 & 73.6 &   0.9 & 35.215072 &   0.000001 & 1.895 & 0.004 & 0.574 & 0.001 & 0.1249 & 0.0002 \\
8854.634287 &   9.0 &  0.6 & 6094.6 &    8.6 & 64.1 &   0.9 & 35.308292 &   0.000001 & 1.762 & 0.003 & 0.567 & 0.001 & 0.1284 & 0.0002 \\
8861.624299 &  16.1 &  0.8 & 6117.1 &    8.7 & 71.8 &   1.1 & 35.214263 &   0.000001 & 1.806 & 0.005 & 0.545 & 0.001 & 0.1241 & 0.0002 \\
8861.707136 &  14.6 &  0.7 & 6113.0 &    8.6 & 70.8 &   0.9 & 35.198146 &   0.000001 & 1.795 & 0.004 & 0.537 & 0.001 & 0.1227 & 0.0002 \\
8883.567735 &  15.6 &  1.0 & 6088.7 &    8.6 & 75.0 &   1.7 & 35.455835 &   0.000001 & 1.721 & 0.009 & 0.533 & 0.001 & 0.1215 & 0.0004 \\
8887.493210 &  11.7 &  0.6 & 6098.0 &    8.6 & 67.3 &   0.9 & 35.326083 &   0.000001 & 1.735 & 0.003 & 0.528 & 0.001 & 0.1214 & 0.0002 \\
8887.596530 &  13.0 &  0.7 & 6097.9 &    8.6 & 67.8 &   0.9 & 35.313566 &   0.000001 & 1.750 & 0.003 & 0.538 & 0.001 & 0.1227 & 0.0002 \\
8899.553769 &  11.8 &  0.7 & 6093.7 &    8.6 & 72.3 &   1.0 & 35.331840 &   0.000001 & 1.725 & 0.004 & 0.533 & 0.001 & 0.1210 & 0.0002 \\
8908.534212 &  15.5 &  0.6 & 6117.8 &    8.7 & 64.7 &   0.9 & 35.174930 &   0.000001 & 1.829 & 0.004 & 0.551 & 0.001 & 0.1219 & 0.0002 \\
8908.651325 &  16.2 &  0.6 & 6119.5 &    8.7 & 67.8 &   1.1 & 35.199129 &   0.000001 & 1.791 & 0.005 & 0.548 & 0.001 & 0.1225 & 0.0002 \\
8910.484569 &  13.5 &  0.6 & 6121.2 &    8.7 & 72.1 &   0.9 & 35.157370 &   0.000001 & 1.795 & 0.004 & 0.557 & 0.001 & 0.1239 & 0.0002 \\
8915.440466 &  13.7 &  0.6 & 6096.9 &    8.6 & 74.5 &   0.8 & 35.285193 &   0.000001 & 1.763 & 0.003 & 0.543 & 0.001 & 0.1211 & 0.0002 \\
8915.533771 &  11.6 &  0.5 & 6088.2 &    8.6 & 77.4 &   1.2 & 35.348770 &   0.000001 & 1.721 & 0.006 & 0.533 & 0.001 & 0.1212 & 0.0003 \\
9656.565329 &  16.8 &  0.9 & 6063.3 &    8.6 & 62.0 &   1.3 & 35.432578 &   0.000001 & 1.693 & 0.007 & 0.529 & 0.001 & 0.1235 & 0.0003 \\
9658.443411 &  18.9 &  0.8 & 6079.5 &    8.6 & 65.7 &   1.3 & 35.369161 &   0.000001 & 1.702 & 0.008 & 0.529 & 0.001 & 0.1241 & 0.0003 \\
9659.567838 &  16.5 &  0.8 & 6078.8 &    8.6 & 68.7 &   0.8 & 35.317963 &   0.000001 & 1.740 & 0.004 & 0.543 & 0.001 & 0.1243 & 0.0002 \\
9670.406369 &  13.2 &  0.7 & 6067.5 &    8.6 & 68.3 &   0.9 & 35.399315 &   0.000001 & 1.742 & 0.004 & 0.544 & 0.001 & 0.1229 & 0.0002 \\
\bottomrule
\caption{HD79210 radial velocities and activity indicators. RVs as calculated using TERRA, and activity indicators as calculated by the HARPS-N DRS 3.7.}
\end{longtable}
    \clearpage

\bibliography{ref.bib}


\clearpage



\clearpage






\end{document}